\documentclass{elsart}
\usepackage{epsfig}
\usepackage{graphicx}
\usepackage{lscape}
\usepackage{subfigure}
\usepackage{setspace}
\journal{Chinese Physics B}
\begin{document}

\begin{frontmatter}

\title{ Kinetic Blume-Capel Model with Random Diluted Single-ion Anisotropy  }
\author{Gul Gulpinar\corauthref{cor}$^{a}$},
\corauth[cor]{corresponding author.}\ead{gul.gulpinar@deu.edu.tr}
\author{Erol Vatansever$^{a}$}

\address{$^{a}$Dokuz Eyl\"{u}l University, Department of Physics, 35160-Buca, \.{I}zmir, Turkey }

\date{\today}

\begin{abstract}
\indent This investigation employs a square lattice and Glauber dynamics methodology
to probe the effects  of diluteness in the crystal-field interaction in a Blume-Capel Ising system under an oscillating magnetic-field. Fourteen different phase diagrams have been observed  in temperature-magnetic-field space as the concentration of the crystal-field interactions is varied. Besides, a comparison is given with the results of the pure spin-1 Ising systems.
\end{abstract}
\begin{keyword}
 Quenched disorder, Random crystal field, Dynamical critical points, Kinetic Blume-Capel Model.
 \PACS 0550, 0570F, 6460H, 7510H
\end{keyword}
\end{frontmatter}

\section{Introduction}
\label{Intro}
\indent  Ising spin-1 systems, with density as an added degree of freedom,
have been utilizied  to investigate  a diverse range of
systems: materials with mobile defects, structural glasses \cite{Kirkpatrick},
the superfluid transition in $He^{3}-He^{4}$ mixtures  \cite{Blume}, frustrated
Ising lattice gas systems \cite {Nicodemiarenzon}, binary
fluids, binary alloys and frustrated percolation \cite {Coniglio}. Recently,
Blume-Capel (BC) \cite{Blumecapel} and Blume-Emery-Griffiths (BEG) models  have been used
to probe the phenomenon of inverse melting, a phenomena observed in diverse class of systems such
as  colloids, polymers, miscelles, etc \cite{Schupperangelini}. \\
\indent On the other hand, it is well known that the effects upon criticality and
resulting phase diagrams, due to underlying competing interactions in various
 spin-1 Ising systems can be complicated. Since then, many previous investigations have been focused on various  novel types of competing interactions have been the focus of previous studies using the BEG  model in conjunction with renormalization-group \cite{WortismcKay1,Branco1,Branco2,Snowmant}
  and/or mean-field methodologies \cite{McKay2,Hoston,Sellitto}.
Besides, effects of disorder on magnetic systems have
been systematically studied, not only for theoretical interests
but also for the identifications with experimental realizations \cite{Bouchiat,Katsumata,Belanger}.
It has been shown by renormalization group arguments that first-order transitions are  replaced by continuous transition, consequently  tricritical points and critical end points are depressed in temperature, and a finite amount of disorder will suppress them \cite{huifaliong}. \\
\indent An special magnetic system with disorder
is spin-1 Ising model with quenched
diluted single ion anisotropy is
used to model phase separations of superfluidity
for helium mixtures in aerogel \cite{Maritan,Buzano}.
Due to this fact, various researchers have been motivated to study the effect of the crystal field disorder on the multicritical phase diagram of BC model via effective field theory \cite{kaneyoshi} and mean field approach \cite{Yusuf,ezbocccarnei},cluster variation method \cite{buzano}, as well as  by introducing an external random field \cite{kauffman}. Whereas, Branco et al. considered the effects of random crystal
fields using real-space RG \cite{Branco1,Branco2}
and mean-field approximations
\cite{Branco2,Branco3} for both BC and BEG model Hamiltonians, respectively.
Recently, Snowman  has employed a hierarchical lattice and renormalization-group methodology to probe the effects
of competing crystal-field interactions in a BC model \cite{Snowman4} .
Finally, Salmon and Tapia have studied the  multicritical behavior of the BC model with infinite-range interactions  by introducing quenched disorder in the crystal field $\Delta_{i}$, which is represented by a superposition of two Gaussian distributions \cite{tapiasalmon}. \\
\indent While the equilibrium properties of the BC model with random single ion anisotropy have been studied extensively, as far as we know, the kinetic aspects of the model
have not been investigated via Glauber dynamics.
Therefore, the purpose of the present
paper is, to present a study
of the kinetics of the spin-1 BC model
with a quenched two valued random crystal field in the presence of a time-dependent oscillating external
magnetic field. We make use of  Glauber-type stochastic dynamics to represent the time evolution of the system  \cite{Glauber}.
More precisely, we have
obtained the  dynamic phase transition (DPT)
points and presented phase diagrams in constant crystal field and  the reduced magnetic field amplitude versus reduced temperature plane for various values of the crystal field concentration.
This type of calculation
for pure BC model, was first performed by
Buendia and Machado \cite{Buendia}. They have presented only two phase diagrams in the temperature-magnetic field plane for the pure spin-1 BC model. Later, Keskin et. al. have shown that one of the two phase diagrams in Ref \cite{Buendia} was incomplete; i.e., they had missed a very important part of the phase diagram due to the reason that they did not make the calculations for higher values of the amplitudes of the external oscillating magnetic field \cite{KeskinPrevE72}. Keskin et. al. presented the phase diagrams in the reduced magnetic field amplitude (h) and reduced temperature (T) plane and calculated five distinct phase diagram topologies.
Recently,  Elyadari et.al. has investigated the kinetic Blume Capel Model with
a random crystal field distributed according to the following law: $P(\Delta_{i})=\displaystyle p\delta(\Delta_{i}-\Delta(1+\alpha))+(1-p)\delta(\Delta_{i}-\Delta(1-\alpha))\ $. This  kind of  random crystal field has been
introduced to study the critical behavior of $^{3}He-^{4}He$ mixtures in random media (aerogel)
modeled by the spin-1 Blume-Capel model \cite{Yadari}. In their model, the negative crystal field value corresponds to the field at the
pore-grain interface and the positive one is a bulk field that controls the concentration of $^{3}He$ atoms.
In other words, in this kind of  randomness the crystal field value is finite for each site about its amplitude takes one of the values
$\Delta(1+\alpha)$ or $\Delta(1-\alpha)$ with equal and fixed probabilities ($p=1/2$).
While in our current investigation, we have focused on another kind of randomness in which one can see the effect of vacancy  of the single-ion anisotropy on the dynamical phase diagrams of the kinetic Blume-Capel model.
Previous equilibrium studies has revealed that this kind of randomness
leads to a rich variety of phase diagrams with type being according to
the concentration $p$ of active local crystal fields. \cite{kaneyoshi,Yusuf}.
With this motivation we have performed numerical calculations for various values of the crystal field concentrations in order to observe the effect of the quenched vacancy in the crystal field on the five different kinetic phase diagram topologies found by Keskin and co-workers \cite{KeskinPrevE72}. \\
\indent Meanwhile, it is worthwhile to stress that the DPT was first found
in the study of the kinetic Ising system in an oscillating field \cite{Tome}, and it was followed by Monte Carlo simulation researches of kinetic Ising models \cite{Acharyya,Chakrabarti}.
Further,
Tutu and Fujiwara \cite{Tutu} represented a systematic method for
obtaining the phase diagrams in DPTs, and constructed a
general theory of DPTs near the transition point based on a
mean-field description, such as Landau$ '$s general treatment
of the equilibrium phase transitions. DPT may also have
been observed experimentally in ultrathin Co films on
Cu(001) \cite{Jiang} by means of the surface magneto-optic Kerr effect and in ferroic systems (ferromagnets, ferroelectrics and ferroelastics) with pinned domain walls \cite{Kleemann} and ultra-thin $[Co/Pt]_{3}$ multilayer \cite{Robb}.
In addition, reviews of earlier research on the DPT and
related phenomena are found in Ref \cite{Chakrabarti}.\\
\indent
The paper is organized as follows: In Sec.2, we discuss the kinetic BC model with single ion isotropy briefly.
Moreover, the derivation of the mean-field dynamic equations of motion is
given by using a master equation formalism in the
presence of an  oscillating external magnetic
field is also given in Sec.2. In Sec.3, the DPT points are calculated and the phase
diagrams presented. Finally, Sec.4  represents the summary and
conclusions.

\section{DYNAMICAL EQUATIONS FOR THE MEAN VALUES}
The generalization of the kinetic BC model for a quenched random crystal field is given by the Hamiltonian,
\begin{equation} \label{1}
\displaystyle  \hat{H}=\displaystyle -J \sum_{\langle ij \rangle }S_{i}S_{j}-\sum_{\{i\}}\Delta_{i}S_{i}^{2}-H\sum_{\{i\}}S_{i}\;,
\end{equation}
where the spin variables $S_{i}=0,\pm1$ on a square lattice. The first and the
second sums are over nearest neighbor pairs. The exchange interaction
with strength $J>0$ is responsible for the ferromagnetic ordering,
while the random single ion anisotropy $\Delta_{i}$ is given by the following
joint probability density:
\begin{equation} \label{2}
P(\Delta_{i})=\displaystyle p\delta(\Delta_{i}-\Delta)+(1-p)\delta(\Delta_{i})\;.
\end{equation}

Finally, $H $ is a time-dependent external oscillating magnetic field and  given by,
\begin{equation}\label{3}
H(t)=H_{0}cos(\omega t)\;,
\end{equation}
here $H_{0}$ and $\omega=2\pi\nu$ denote the amplitude and the angular
frequency of the oscillating field respectively. \\
\indent When we put this system
in contact with a heat reservoir at temperature T, the spin variables $S_{i}$ can be
considered as stochastic functions of time. The system evolves according to a Glauber-type stochastic
process at a rate of $\frac{1}{\tau}$ transitions per unit time.
More precisely, we will follow the heat-bath prescription \cite{Janke}:
the new value of the spin variable at site $i$ ($S_{i \; new}$) is determined by testing all its possible states in the heat-bath
of its (fixed) neighbors (here four on a square lattice):
\begin{equation} \label{4}
w_{i}(S_{i\; old}\rightarrow S_{i\; new})=\frac{1}{\tau}\frac{\exp\{-\beta \Delta E(S_{i \; old}\rightarrow  S_{i\; new} )\}}{\sum \exp\{-\beta \Delta E(S_{i \; old} \rightarrow S_{i\; new})\}}\;,
\end{equation}
where $ \beta=\frac{1}{kT}$ and $\tau$  defines a time scale (characteristic mean time interval for one spin flip),
and
\begin{equation}\label{5}
\Delta E (S_{i\; old}\rightarrow S_{i\; new})=(S_{i\; old}- S_{i\; new})\left( J\sum_{\langle j \rangle}S_{j}+H\right)-\left(S^{2}_{i\; old}- S^{2}_{i\; new}\right)\Delta_{i} \; ,
\end{equation}
give the changes in the energy of the system in the case of flipping of the
$i^{th}$ spin in the lattice. If we define $P(S_{1} ,S_{2} , . . . ,S_{N}; t)$ as the probability that the system
has the configuration $\{S_{1},S_{2} , . . . ,S_{N}\}$, at time t. Making use of master equation formalism \cite{Glauber}, one can write the time derivative
of $P(S_{1} ,S_{2} , . . . ,S_{N}; t)$ as,
\begin{equation}\label{6}
\begin{array}{ccc}
\displaystyle \frac{d}{dt}P(S_{1},...,S_{N};t) & = &-\sum_{i} \sum_{S_{i \;old}\neq S_{i \;new}} w_{i}(S_{i \; old} \rightarrow S_{i\;new})P(S_{1},...,S_{i \;old},...S_{N};t) \\
   &  & +\sum_{i}\sum_{S_{i \;old}\neq S_{i \;new}}w_{i}(S_{i\; new}\rightarrow S_{i\; old})P(S_{1},...,S_{i \;new},...S_{N};t)\;.
 \end{array}
\end{equation}

The detailed balance condition reads,
\begin{equation}\label{7}
\frac{w_{i}(S_{i\; old}\rightarrow S_{i\; new})}{w_{i}(S_{i\; new}\rightarrow S_{i\; old})}=\frac{P(S_{1} ,S_{2} , . . . ,S_{i\; new},...,S_{N})}{P(S_{1} ,S_{2} , . . .,S_{i\; old},...S_{N})}\;.
\end{equation}

  In addition, substituting the possible values of $S_{i \;new}$ and  $S_{i \;old}$ , one obtains:
\begin{equation}\label{8}
\begin{array}{ccc}
\displaystyle w_{i}(1 \rightarrow 0)=\displaystyle w_{i}(-1 \rightarrow 0)& = & \displaystyle \frac{1}{\tau}\frac{exp(-\beta \Delta_{i})}{2cosh(\beta \delta)+exp(-\beta \Delta_{i})}\;, \\
\displaystyle w_{i}(1 \rightarrow -1)=w_{i}(0 \rightarrow -1)& = & \displaystyle \frac{1}{\tau}\frac{exp(-\beta \delta)}{2cosh(\beta \delta)+exp(-\beta \Delta_{i})}\;,
 \\
\displaystyle  w_{i}(0 \rightarrow 1)=w_{i}(-1 \rightarrow 1)& = & \displaystyle \frac{1}{\tau}\frac{exp(\beta \delta)}{2cosh(\beta \delta)+exp(-\beta \Delta_{i})}\;,
\end{array}
\end{equation}
where $\delta= J\sum_{\langle j \rangle}S_{j}+H $. At this point one can notice that
 $ w_{i}(S_{i\; old}\rightarrow S_{i\; new})$ does not
depend on the value $S_{i\; old}$, we can write $ w_{i}(S_{i\; old}\rightarrow S_{i\; new})=w_{i}(S_{i\; new})$,
then the master equation becomes:
\begin{equation}\label{9}
\begin{array}{ccc}
 \displaystyle \frac{d}{dt}P(S_{1},...,S_{N};t)&=&-\sum_{i}\sum_{S_{i\;old}\neq S_{i\;new}} w_{i}(S_{i\;new})P(S_{1},...,S_{i \;old},...S_{N};t)\\
   &  &+\sum_{i}w_{i}(S_{i\; old}) \sum_{S_{i \;old}\neq S_{i \;new}}P(S_{1},...,S_{i \;new},...S_{N};t) \;.
\end{array}
\end{equation}

On the other hand, the sum of probabilities is normalized to one so that  by
multiplying both sides of Eq.(\ref {9}) by $S_{p}$ and taking the average,
one obtains,
\begin{equation}\label{10}
\begin{array}{ccc}
\tau\frac{d}{dt}\langle S_{p} \rangle & = &\displaystyle - \left \langle S_{p} \right \rangle+ \left \langle \int P(\Delta_{i}) \frac{2\sinh\left(\beta\left[ J\sum_{\langle j \rangle}S_{j}+H\right]\right)}{2cosh(\beta \left[ J\sum_{\langle j \rangle}S_{j}+H\right])+exp(-\beta \Delta_{i})} d\Delta_{i}\right \rangle\;.
\end{array}
\end{equation}

Finally, after integration over the distribution of $P(\Delta_{i})$ and making use of mean field approximation,
 the kinetic equation of the magnetization becomes,
\begin{equation}\label{11}
\begin{array}{ccc}
\Omega \displaystyle\frac{d}{d\xi}m &= & \displaystyle -m+p\frac{2sinh(\frac{m+hcos(\xi)}{T})}{2cosh(\frac{m+hcos(\xi)}{T})+exp(-\frac{d}{T})}
 +(1-p)\frac{2sinh(\frac{m+hcos(\xi)}{T})}{2cosh(\frac{m+hcos(\xi)}{T})+1}\;,
\end{array}
\end{equation}
where $\xi=\omega t,\; m=\langle S \rangle,\; T=(\beta z J)^{-1},\; d=\Delta/zJ,\; h=H_{0}/zJ\; $ .
In these equations the variable $\Omega$ was defined as the ratio between the external field frequency $\omega$ and the frequency of spin flipping ($f=1/\tau$), i.e., $\Omega=\omega \tau=\omega/ f$.
Here we consider a cooperatively interacting many-body system, driven by an oscillating external perturbation, an oscillating magnetic field so that  the thermodynamic response of the system, the magnetization, will then also oscillate with necessary modifications in its form \cite{Chakrabarti}.
Moreover, the time dependence of magnetization can be one of two
types according to whether they have or do not have the
property:
\begin{equation}\label{12}
\displaystyle m(\xi+\pi)=\displaystyle-m(\xi)\; .
\end{equation}
A solution that satisfies Eq.(\ref{12}) is called symmetric solution;
it corresponds to a paramagnetic (P) phase. In this
solution, the magnetization $m(\zeta)$ oscillates around the
zero value and is delayed with respect to the external
field. Solutions of the second type, which do not satisfy
Eq.(\ref{12}), are called nonsymmetric solutions; they correspond to
a ferromagnetic (F) phase. In this case, the magnetization
does not follow the external magnetic field any
more, but, instead, oscillates around a nonzero value. Eq. (\ref{11}) is solved numerically by using fourth order Runge-Kutta method for fixed values of $T,d,p$, and $\Omega$. Throughout this study we have fixed $\Omega=2\pi$, $J=1$ and $z=4$ for a
given set of parameters and initial values. The results
are presented in Figs.1(a)-(c). Here, we can see three different
solutions: F,P and coexistence of F and P (F+P).
In Fig.1(a), only the symmetric solution is always obtained, and, hence,
we have a paramagnetic (P) solution; but, in Fig.1(b),
only the nonsymmetric solutions are found, and we,
therefore, have a ferromagnetic (F) solution. One can observe from these figures that these
solutions do not depend on the initial values. On the
other hand, in Fig.1(c), both the F and P phases
exist in the system this case corresponds to the coexistence
solution (F + P). As can be seen in Fig.1(c) explicitly, the solutions depend on
the initial values.
\section{DYNAMIC PHASE TRANSITION POINTS AND PHASE
DIAGRAMS}
In order to obtain the dynamic phase boundaries between
three phases or regions in that are given Figs.1(a)-(c),
one should calculate
the DPT points. The DPT points are
obtained by investigating the behavior of the average
magnetization in a period as a function of the reduced
temperature.
\begin{equation}  \label{13}
M=\displaystyle\frac{1}{2\pi}\int\limits_{\xi_{0}}^{\xi_{0}+2\pi}m(\xi)d\xi,
\end{equation}

Here  $m(\xi)$ is a stable and  periodical function. In general our solution stabilizes after $6000$ periods. In this manner, $\xi_{0}$ can take any  value after this transient.
In the high field  and high temperature region
time dependent staggered magnetization  follows the reduced
external magnetic field  within a single period
which corresponds to vanishing  time average
of the dynamical order parameter (paramagnetic phase).
Whereas, at low field values the magnetization can not fully switch sign in a
single period and the time average of the magnetization in a period
is non zero and  consequently  ordered or ferromagnetic phase  arises.
Fig.2 represents the reduced temperature dependence of the
average magnetization (M) for various values of magnetic field amplitude h and crystal field concentration (p) while $d=-0.25$. In these figures arrows denote the transition temperatures.
 In Fig.2(a), we give the case for $h=0.70,p=0.50$.  In this
case, the system represents re-entrance with two sequential first order phase transitions which take
place at $T_{t1}$ and $T_{t2}$. While Fig.2(b) exhibits the reduced temperature dependence of the dynamical order parameter for $h=0.70,p=0.75$. For these values of the parameters, BC model with random single ion anisotropy
undergoes a first and a second phase transition sequentially. In Fig.2(c) we give an example of second order phase transition from ordered to disordered phase for $h=0.4,p=0.75$. Eventually, Fig.2(c) illustrates an first order
phase transition which occurs for $h=0.75,p=0.75$.\\
\indent
Fig.3(a) illustrates the thermal variations of M  for various values of crystal field concentration (p)
for vanishing external field. The number accompanying each curve illustrates
the value of p. Here the crystal field has negative sign ($d=-0.3$) since then the critical temperature increases with increasing diluteness for fixed h.\\
 \indent On the other hand, it is well known fact that in the static limit ($\omega=0.0$) the dynamic transition disappears and the phase boundary in the $h-T$ plane collapses to a line with $h=0$ and ending at $T=T_{c}$ , the  static transition temperature of the unperturbed system \cite{Chakrabarti}. Fig.3(b) shows the thermal variations of M and for several values of static h while $d=-0.5$. In addition, Fig.3(c) gives the behaviors of M and  as a function of static h for $d=-0.5$ and several values of T. One can see from these figures that the system does not  undergo any phase transitions for static h. Consequently, we can conclude that the oscillating external magnetic field induces the phase transitions.\\
 \indent
 We can now focus on the phase diagrams of the system. In Figs.3-5
we represent the calculated mean field dynamical phase diagrams in the (T,h) plane which exhibit the
effect of the randomness in the crystal field on the five different phase diagram topologies
of the pure kinetic BC model \cite{KeskinPrevE72}.
First in Fig.3(a), we give the dynamical phase diagram of the two-dimensional kinetic BC model
with bimodal crystal field distribution for $d=0.25$. The number accompanying each curve denotes
the value of the crystal field concentration (p). The outermost curve corresponds to the pure BC model
with no quenched randomness ($p = 0$).As crystal field quenched randomness is introduced with decreasing values of p, ordered phases and first-order phase transitions recede. This result is consistent with the RG theory predictions given in Ref.\cite{huifaliong}. Finally, we should stress that similar phase diagrams
were also obtained in the kinetic of the mixed spin-$\frac{1}{2}$ and
spin-1 Ising ferromagnetic system \cite{Buendia} as well as the kinetic
spin-$\frac{1}{2}$ Ising model \cite{Tome}. The reason that the phase diagram
is similar to the one obtained for the kinetic spin-$\frac{1}{2}$
Ising model is due to the competition between $J, d$ and $h$. For positive
crystal field values, the Hamiltonian of the spin-1 model gives similar results
to the Hamiltonian of the spin-$\frac{1}{2}$ Ising model. \\
\indent It has been given in Ref.\cite{KeskinPrevE72} that pure kinetic BC model has
four different phase diagram topologies for negative d,
which depend on d values. Now let us discuss the effect of randomness in the single ion isotropy
on these phase diagrams:\\
\indent
(1) For $-0.0104>d\geq-0.4654$, the dynamical phase diagram topology
of the pure kinetic BC model is similar to positive d case but only differs
in that for very low T and h values, one more P+F coexistence region also
exists. The boundary between this F+P region and the F
phase is the first-order line (see  Fig.7(b) in Ref.\cite{KeskinPrevE72}.
In BC model, described by the Hamiltonian given in Eq.(\ref{1}), negative
crystal field interaction ($d=\Delta/zJ<0$) favors the annealed vacancies, namely the
 nonmagnetic states $S_{i}=0$. Fig.3(b) exhibits this fact: with increasing concentration
of negative single ion isotropy ($d=-0.25$) the ordered phase recedes and the tricritical
temperature moves to lower temperatures. Whereas, the coexistence region (F+P) in the low temperature and field region disappears with increasing vacancy in the single ion anisotropy (for $p\leq0.95$).\\
\indent
(2) For $-0.4654< d \leq -0.5543$, the system exhibits two
dynamic tricritical points (DTCP). One of them occurs in similar
places in the phase diagrams for $d=0.25$ and $d=-0.25$ ,
 whereas, the other DTCP occurs in the low
h region. In addition, the first-order phase transition lines exist at the
low reduced temperatures, and h values separate not only the
P+F region from the F phase, but also from the P phase.
When we introduce quench disorder in the crystal field we
found that this topology changes drastically with varying crystal field
concentration ($p$). Our calculations has revealed that
there are four different phase diagram topologies
which depend on $p$ values:\\
\indent(2.a) Type 1 ($p \geq 0.97$): the system has the
same phase diagram with the pure case (see Fig.3(c)).\\
\indent (2.b) Type 2 ($ 0.97 > p \geq 0.85$): The DTCP
in the low temperature and field region disappears. Whereas,
the second order line intersects the $h=0$ axis. Moreover, the
P+F phase recedes and an ordered phase appears in the neighborhood
of $h=0,T=0$ (see Fig.3(d)).\\
\indent (2.c) Type 3 ($ 0.85 > p \geq 0.31$): with increasing single ion isotropy
vacancy, the coexistence region in the low H and low T region moves to higher field
values (see Fig.3(e)). The diagram contains one DTCP first-order phase transition lines merge
and signals the change from a first- to a second-order phase transitions. \\
\indent(2.d) Type 4 ($ 0.31 > p \geq 0.0$): Now the crystal field interaction is rather
diluted. As one can see from Fig.3(f) that the F+P phase coexistence totally disappears
and the phase diagram is similar to the  $-0.0104>d\geq-0.4654$ case.\\
(3)  For $-0.5543< d \leq -0.9891$, pure
system exhibits  an exotic  phase diagram which contains,
three different the F+P regions at low reduced temperatures in addition to the P and
F phases and two dynamical tricritial points.
On the other hand, if one introduces disorder in the crystal field
there are four different phase diagram topologies
which depend on the concentration of the crystal field (see Figs.5(a)-(d). ):\\
\indent(3.a) Type 1 ($ 1.0 \geq p > 0.8$): Although the dynamical phase diagram of the random single ion-anisotropy
BC model has similar topology with the pure kinetic spin-1 BC model, one can observe that (see Fig.7(d) of Ref.\cite{KeskinPrevE72} ) the ordered phase moves to lower temperatures and  the coexistence region in the low temperature and field region shrinks with raising amount of the crystal field randomness and finally disappears at p=0.79.\\
\indent(3.b) Type 2 ($ 0.80 \geq p  > 0.32$):
For  this interval of the crystal field concentration, the ferromagnetic phase expands in the expense of the F+P and P phases  and due to this  effect the first order transition line which takes place in the high field low temperature regime disappears and
the system has only one DTCP.\\
\indent(3.c) Type 3 ($ 0.32 \geq p  > 0.1$): The phase boundary that separates F+P coexistence phases and F phase turns out to be second order as consequence the dynamical TCP moves to zero temperature . Whereas, the boundary between F+P and P phases remains as first order. In addition to the DTCP, system exhibits a dynamical critical end point (DCEP) which is shown Fig. (5.c)\\
\indent(3.d) Type 4 ($ 0.1 \geq p  \geq 0.0$): The behavior differs from Type 3  in the sense that for very
low T and h values, P+F coexistence region  disappears and system represents no DCEP. For this interval of the concentration value, the system has only one DTCP which exists at low temperature and high magnetic field.\\
(4) For $-0.9891>d$, the topology of the phase diagram is dramatically different from the other intervals of the single ion anisotrpy amplitude : it does not include a P+F phase coexistence region at low temperature and low magnetic field for pure spin-1 BC model.
In Figs. 6(a) to (d) we illustrate the four different types of behavior depending on the the p:\\
\indent4(a) Type 1($1 \geq p > 0.994$): For this interval of the crystal field amplitude, the system   exhibits two DTCP's and
the topology of the phase diagram is very similar to the pure case.\\
\indent4(b) Type 2 ($ 0.993 \geq p > 0.14$): The phase boundary that separates F+P coexistence phases and F phase turns to be second order line with decreasing  concentration value of the randomness single-ion anisotropy. As a result of this, the undermost DTCP turns out to be a DCEP.\\
\indent4(c) Type 3 ($ 0.14 \geq p \geq 0$): Finally, for d=-1 and p=0, the system exists one DTCP at low temperature and high magnetic field. The point where  the two boundary lines merge. Also this behavior is similar to kinetic mixed Ising \cite{Buendia} and  kinetic spin-1/2 Ising model \cite{Tome}.


\section{SUMMARY AND CONCLUSIONS}
Within the mean field approach, we have analyzed stationary
states of the spin-1 Blume-Capel model with a random crystal field $\Delta_{i}$ under
a time-dependent oscillating external magnetic field. The time evolution of the system is described
by a stochastic dynamics of the Glauber type.
 We have studied  the time dependence of the magnetization and  the behavior of
the dynamical order parameter as a function of reduced temperature
for reduced magnetic field and  different possibility (p) of the crystal field.
We have  also analyzed thermal variations and temperature dependence of M
for various values of crystal field concentration (p) and for different static reduced magnetic field, respectively.
Moreover, the behavior of M as function of static reduced magnetic field (h) for various values of the reduced temperature have been examined.\\
\indent The dynamic phase transition (DPT) points are found and the phase diagrams are constructed in the reduced magnetic field and temperature plane.
 We have found that the behavior of the system strongly depends on the values of random crystal field or  random single-ion anisotropy. For all (p) and positive values of reduced crystal field (d) the system behaves as the standard kinetic Ising model \cite{Tome}, and also kinetic mixed Ising spin-(1/2,1) model \cite{Buendia}. We have observed that there exist F+P coexistence and first order dynamical phase transitions in the low temperature and high field regime. Whereas, the dynamical phase transitions turns out to be second order with increasing temperature and decreasing magnetic field. Consequently, it  shows that the system exhibits dynamical tricritical point (DTCP). As we have mentioned in detail in the previous section, the introduction of random ion-anisotropy in kinetic spin-1 BC model produces an effect which suppress the F+P coexistence region in the low temperature low field and high field low temperature regimes. We should stress that this result in according with the previous results obtained by Renormalization-Group theory \cite{huifaliong}. Finally we should point out that some of the first-order phase lines and also the dynamic multicritical points (DTCP and DCEP) are very likely artifacts of the mean-field approach.
 The reason of this artifact can be stated as follows:  for  field amplitudes less than the coercive field and temperatures lower than the
 static ferromagnetic - paramagnetic  transition temperature,
 the time dependent  magnetization represents a nonsymmetric stationary solution even zero frequency
 limit. Meanwhile in the absence of the
 fluctuations, the system is trapped in one well of
 the free energy and cannot go to other one \cite{Acharyya}.
 On the other hand, this mean-field dynamic study
 reveals that the random  single ion anisotropy  spin 1 Blume-Capel  Model  represents interesting dynamic phase diagram topologies.
 Since then, we hope that this work can stimulate further studies on kinetic features of
 kinetic random  single ion anisotropy  spin-1 Model systems theoretically and experimentally.
\section{ACKNOWLEDGMENTS}
This research was supported by the Scientific and Technological
Research Council of Turkey (TUBITAK), including
computational support through the TR-Grid
e-Infrastructure Project hosted by ULAKBIM, and by the
Academy of Sciences of Turkey.

\newpage
\begin{figure}[tbp]
\begin{center}
\subfigure[\hspace{0.4cm}]{\label{fig:sub:a}
\includegraphics[width=6.0cm,height=6.0cm,angle=0]{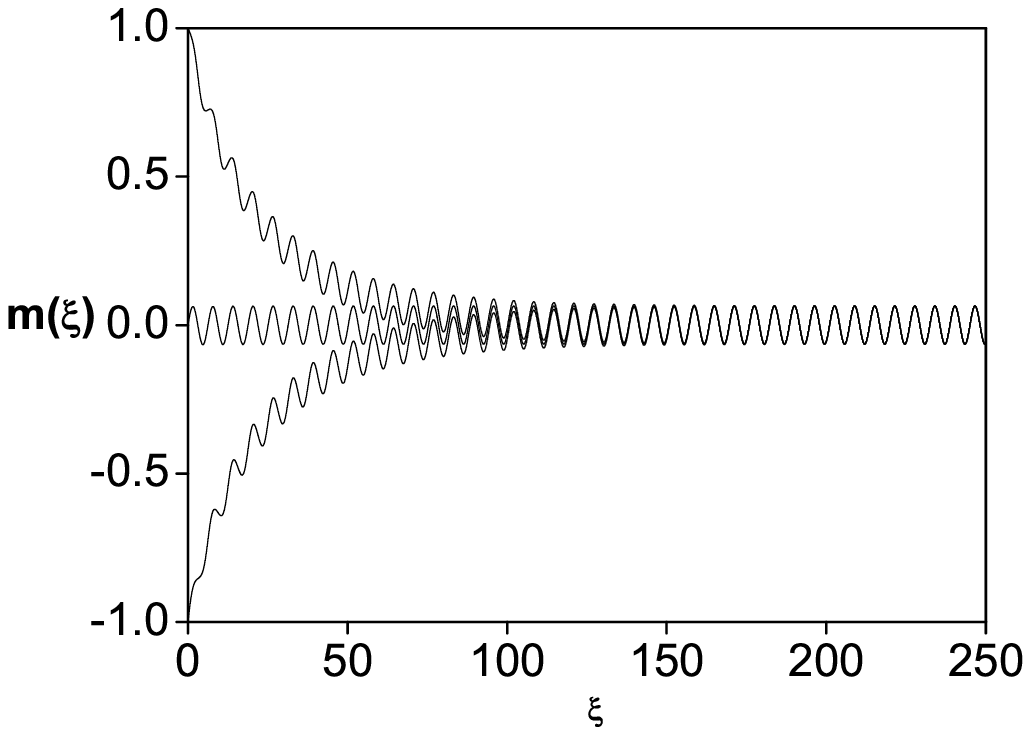}}
\subfigure[\hspace{0.4cm}]{\label{fig:sub:b}
\includegraphics[width=6.0cm,height=6.0cm,angle=0]{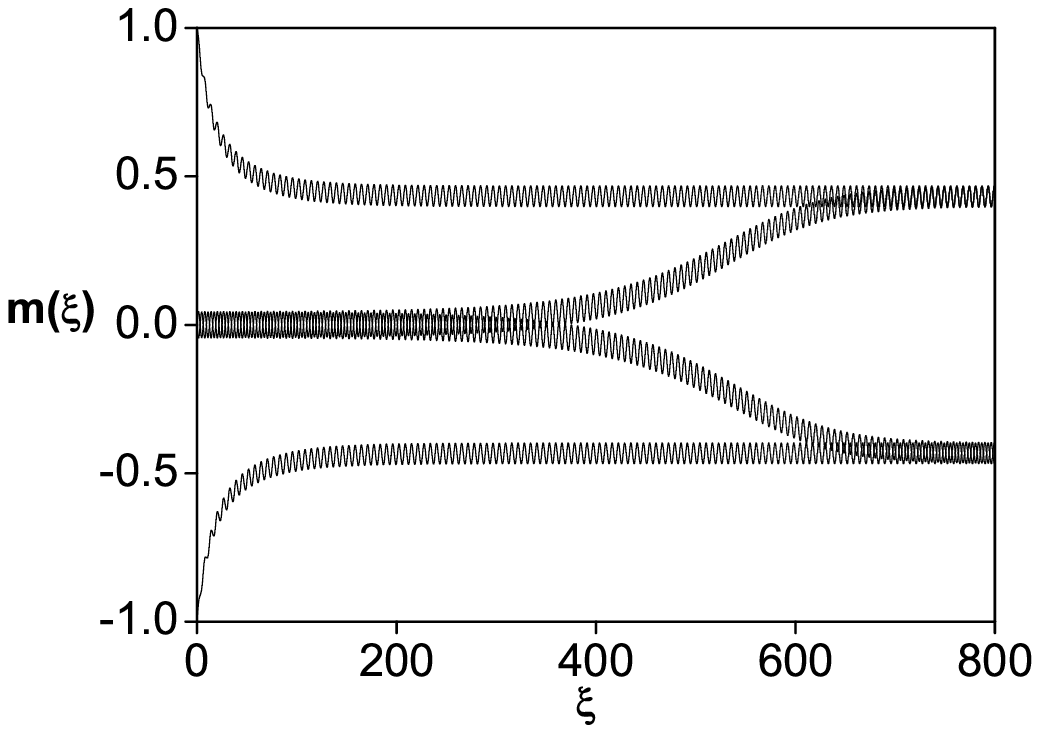}}
\subfigure[\hspace{0.4cm}]{\label{fig:sub:c}
\includegraphics[width=8.0cm,height=6.0cm,angle=0]{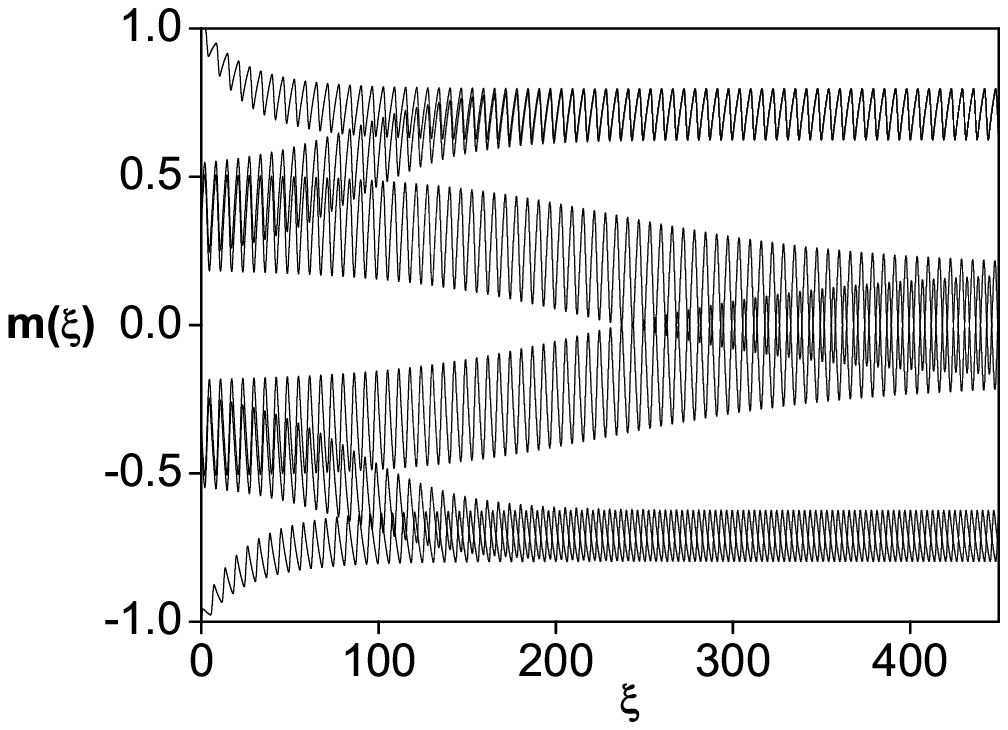}}
\end{center}

\caption{Time variance of the magnetization ($m(\xi)$) while p=0.9 : (a) Corresponding to
a paramagnetic phase (P) for  d=-0.25, h=0.5, and T=0.7; (b) Exhibiting
a ferromagnetic phase (F) for  d=-0.25, h=0.25, and T=0.5; (c)
Representing a coexistence region (F+P) d=-0.25, h=0.75, and T
=0.1.
} \label{fig:sub:Fig2a-Fig2d}
\end{figure}

\begin{figure}[tbp]
\begin{center}
\subfigure[\hspace{0.4cm}]{\label{fig:sub:a}
\includegraphics[width=6.0cm,height=6.0cm,angle=0]{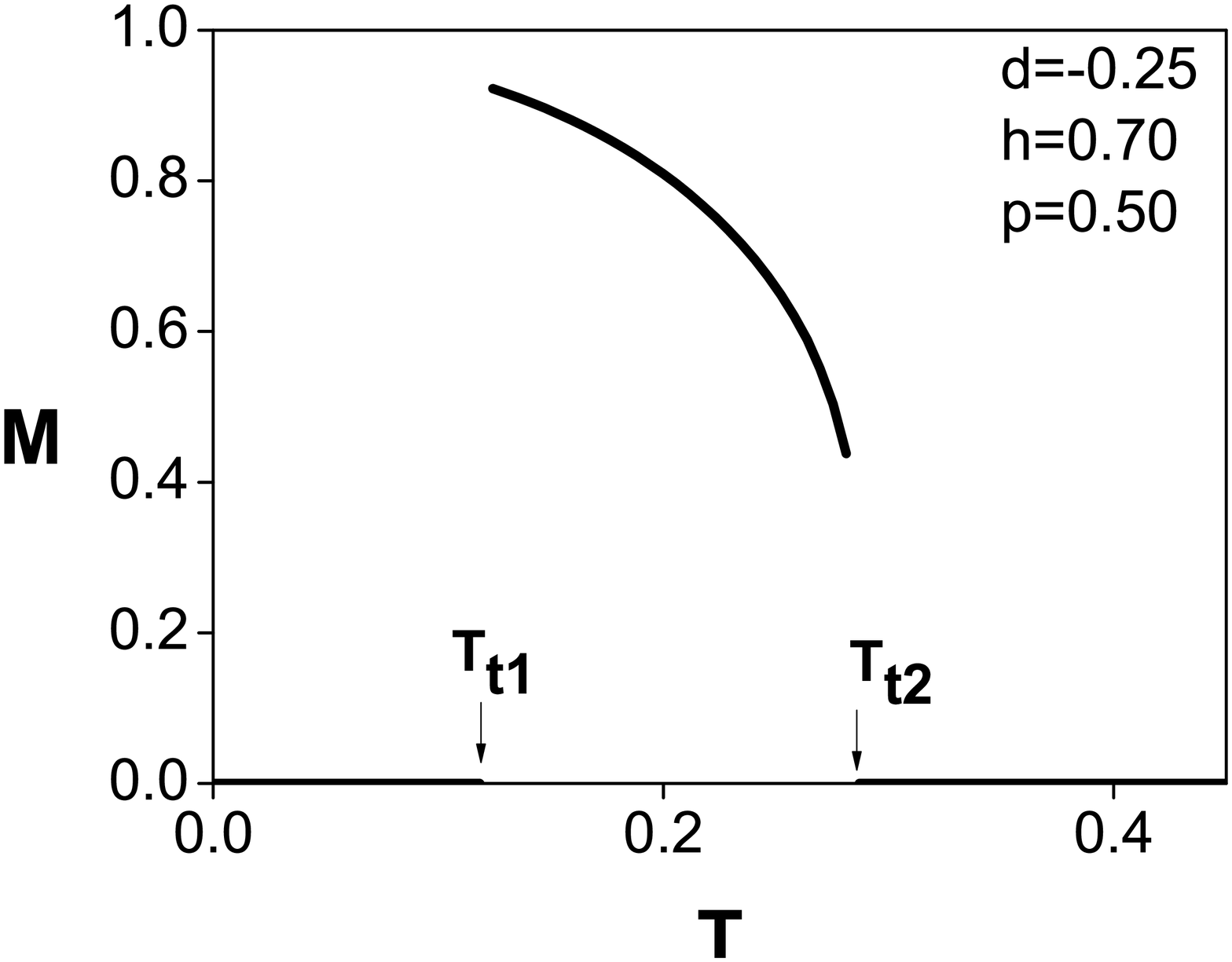}}
\subfigure[\hspace{0.4cm}]{\label{fig:sub:b}
\includegraphics[width=6.0cm,height=6.0cm,angle=0]{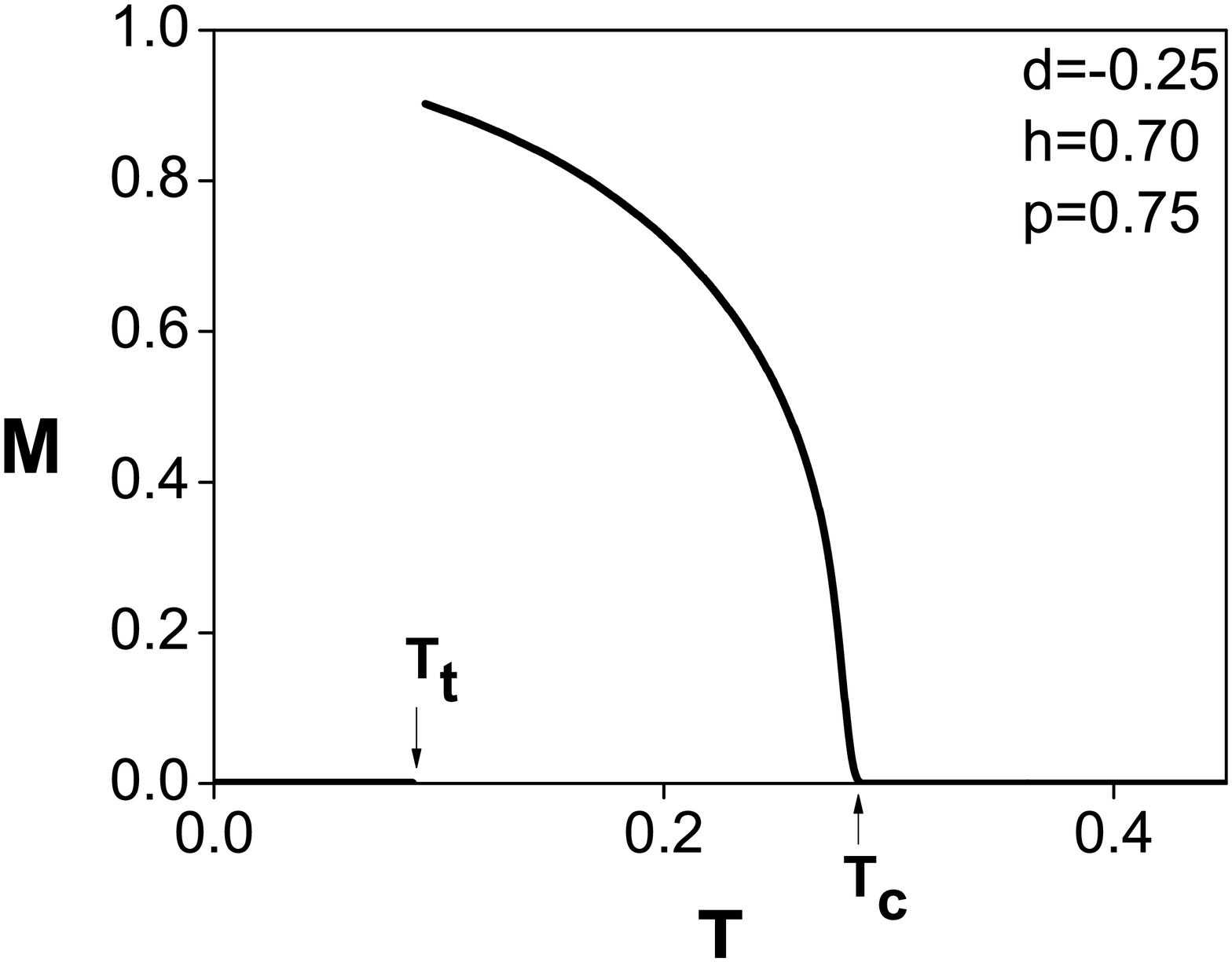}}
\subfigure[\hspace{0.4cm}]{\label{fig:sub:c}
\includegraphics[width=6.0cm,height=6.0cm,angle=0]{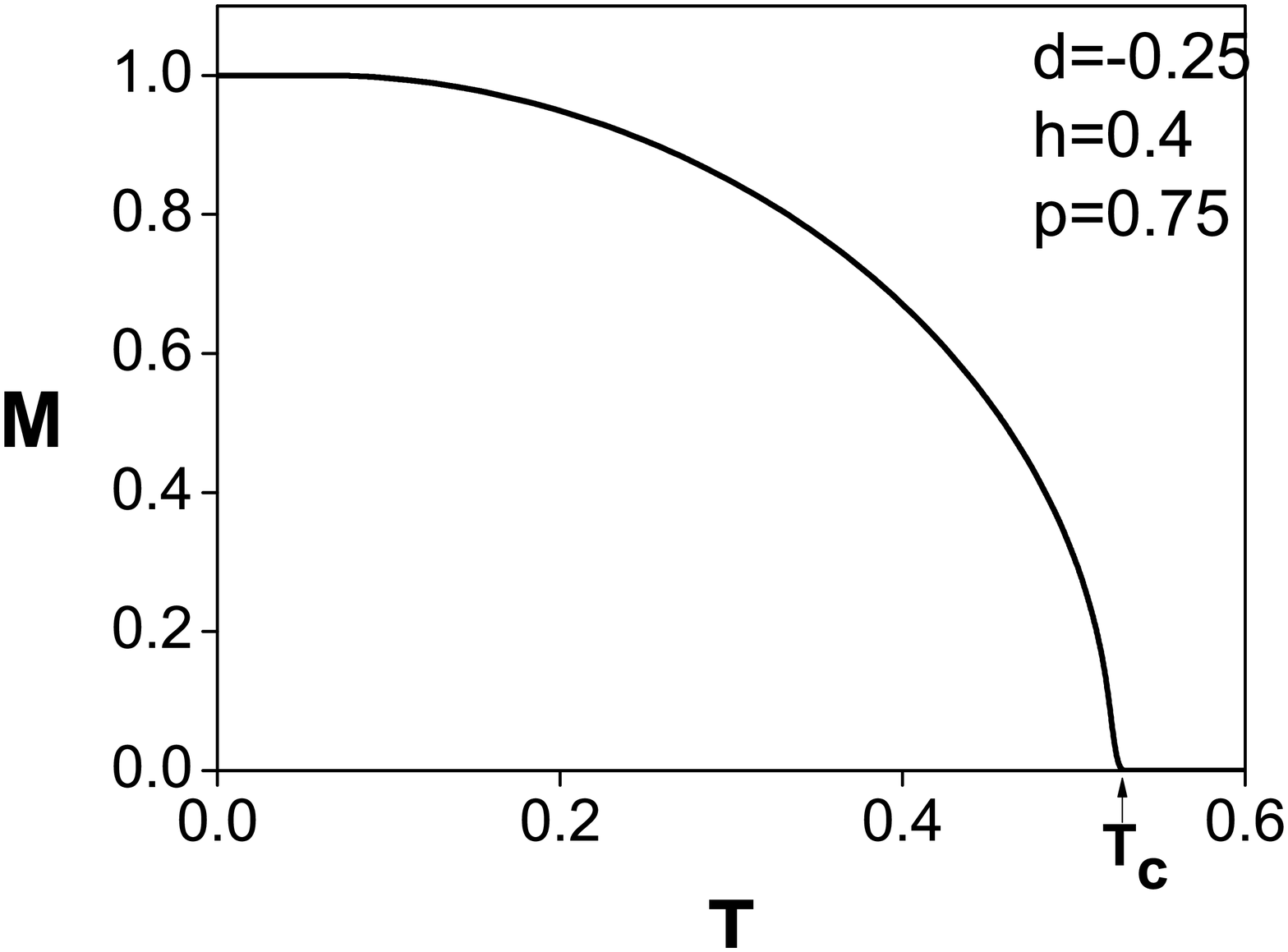}}
\subfigure[\hspace{0.4cm}]{\label{fig:sub:c}
\includegraphics[width=6.0cm,height=6.0cm,angle=0]{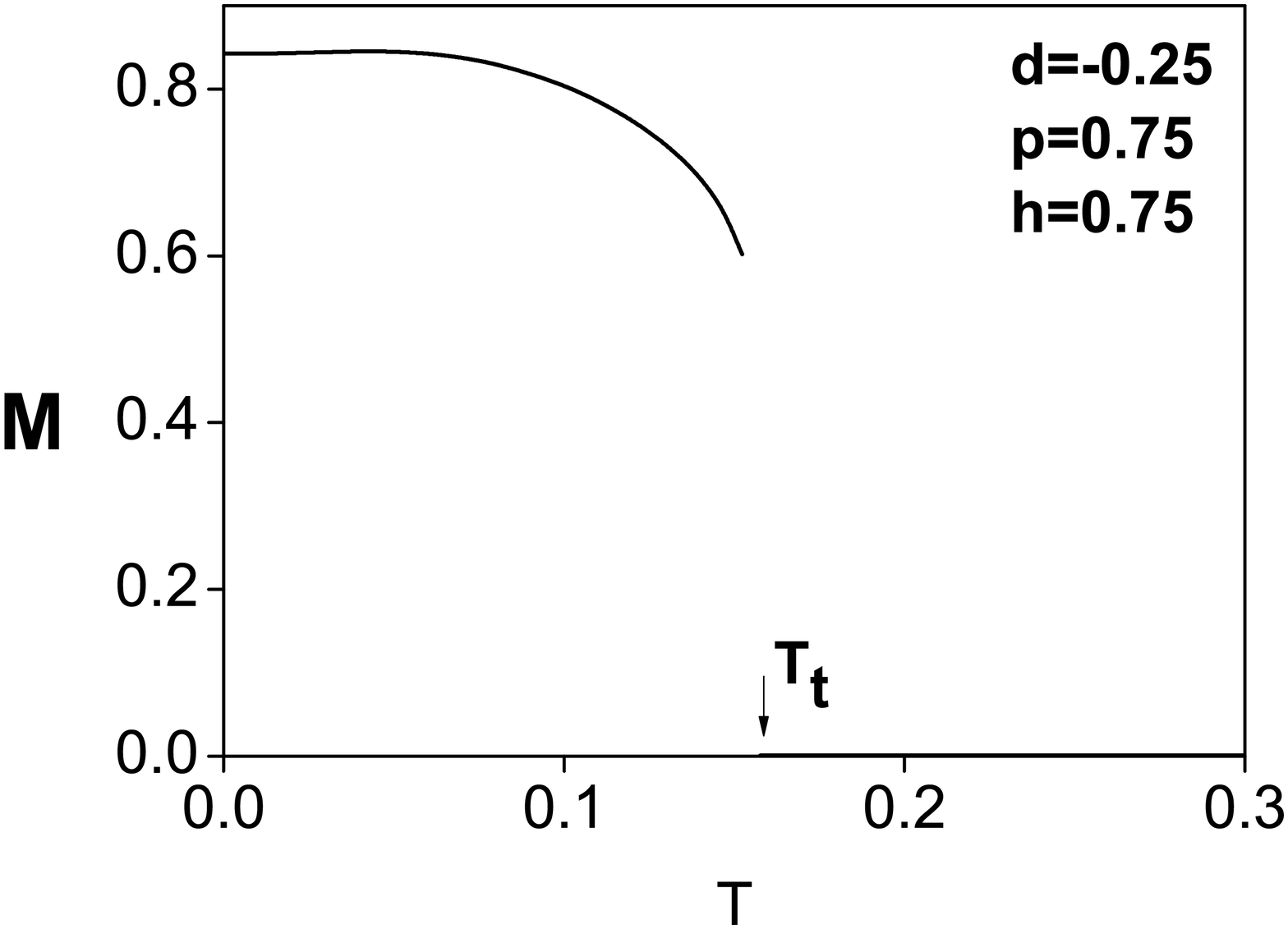}}
\end{center}

\caption{ Dynamical order parameter as a function of
reduced temperature. $T_{c}$ and $T_{t}$ indicate
second and first order phase transition temperatures respectively.  (a) The system under goes two successive first order phase transitions, there exists re-entrance. (b) Two successive phase transitions: the first one is  a first-order and the second one a continuous phase transition and there is re-entrance. (c) The system under goes a second order phase transition. (d) The system shows a first order phase transition.
} \label{fig:sub:Fig2a-Fig2d}
\end{figure}
\begin{figure}[tbp]
\begin{center}
\subfigure[\hspace{0.4cm}]{\label{fig:sub:a}
\includegraphics[width=6.0cm,height=6.0cm,angle=0]{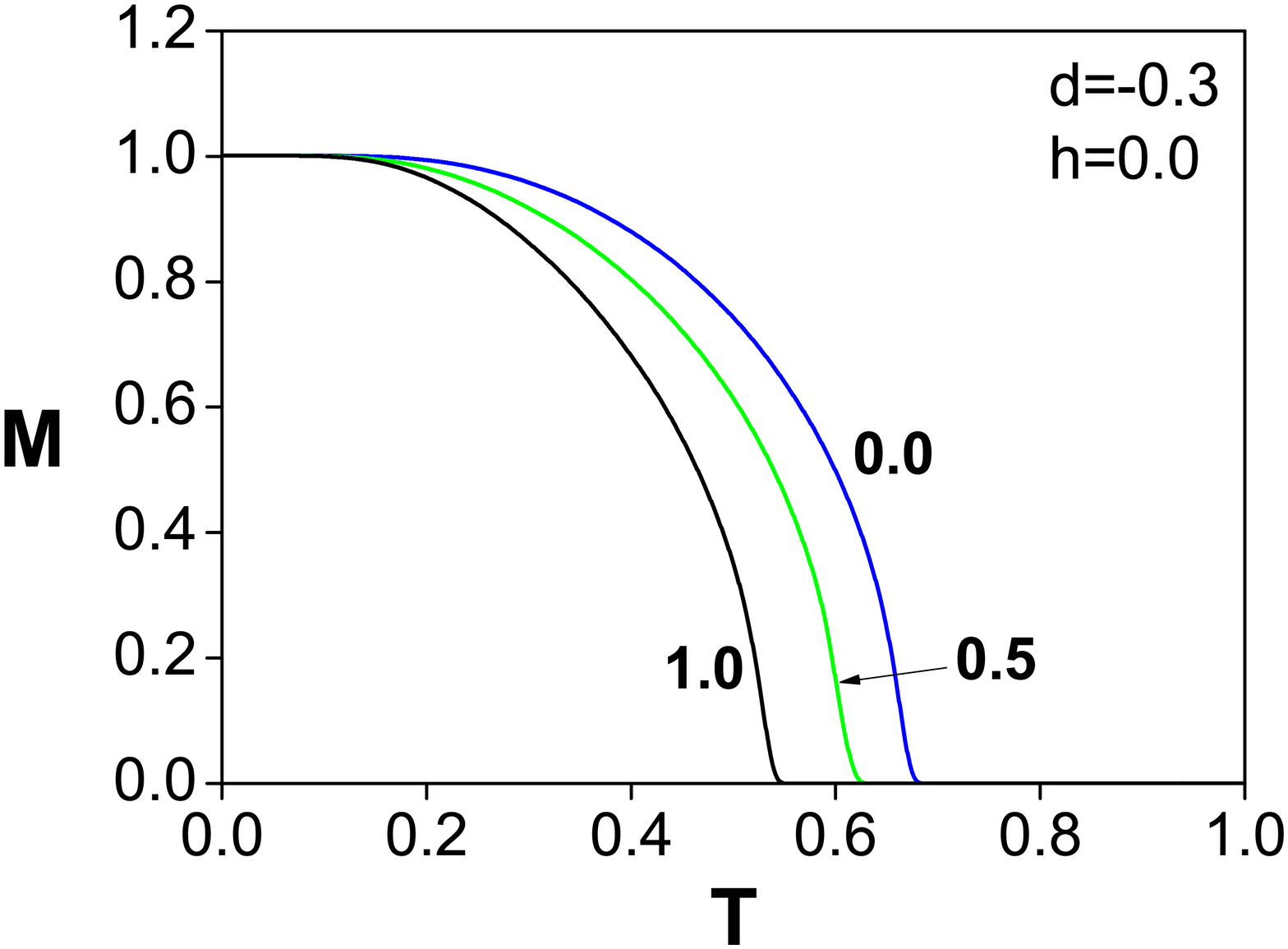}}
\subfigure[\hspace{0.4cm}]{\label{fig:sub:b}
\includegraphics[width=6.0cm,height=6.0cm,angle=0]{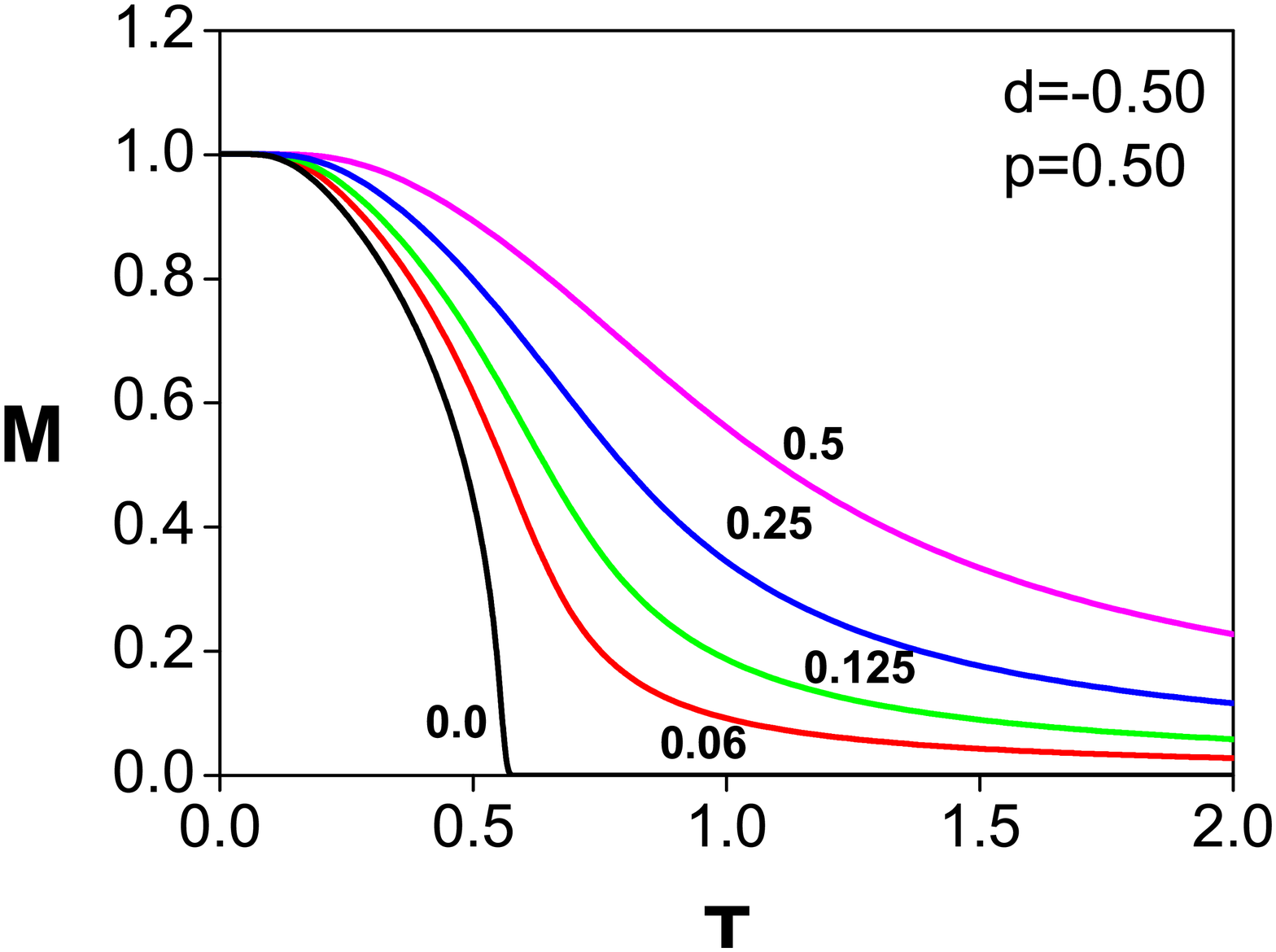}}
\subfigure[\hspace{0.4cm}]{\label{fig:sub:c}
\includegraphics[width=6.0cm,height=6.0cm,angle=0]{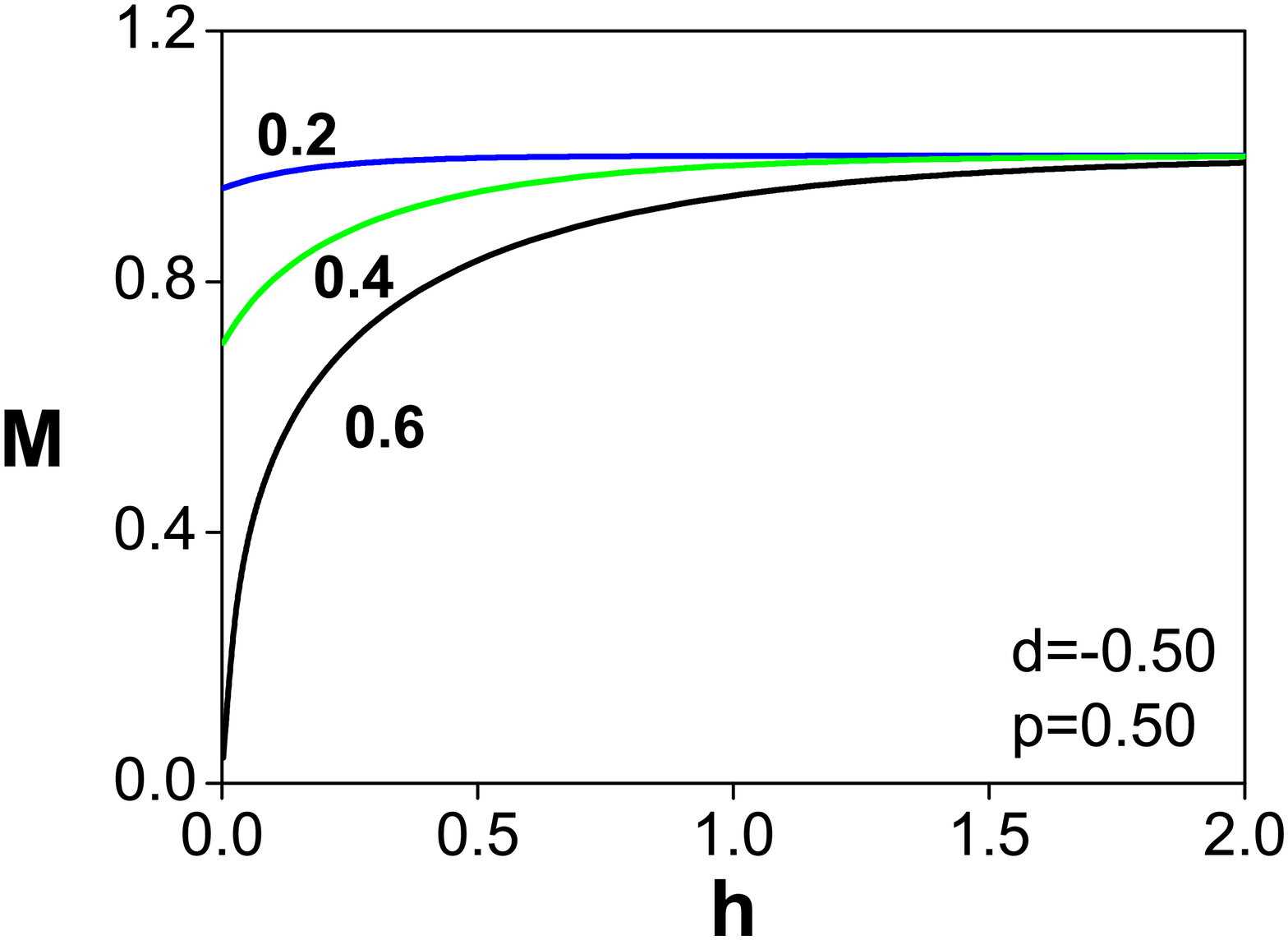}}
\end{center}

\caption{(a) Thermal variations of M  for various values of crystal field concentration (p)
for vanishing external field. The number accompanying each curve illustrates
the value of p. (b) Temperature dependence of M for several values of
static external field amplitudes (h) while p=0.5. The number accompanying each curve
denotes the value of h.  (c) The behavior of M  as function
of static h for d=-0.5. The number accompanying each curve
denotes the value of the reduced temperature (T).} \label{fig:sub:Fig2a-Fig2d}
\end{figure}

\begin{figure}[tbp]
\begin{center}
\subfigure[\hspace{0.4cm}]{\label{fig:sub:a}
\includegraphics[width=6.0cm,height=6.0cm,angle=0]{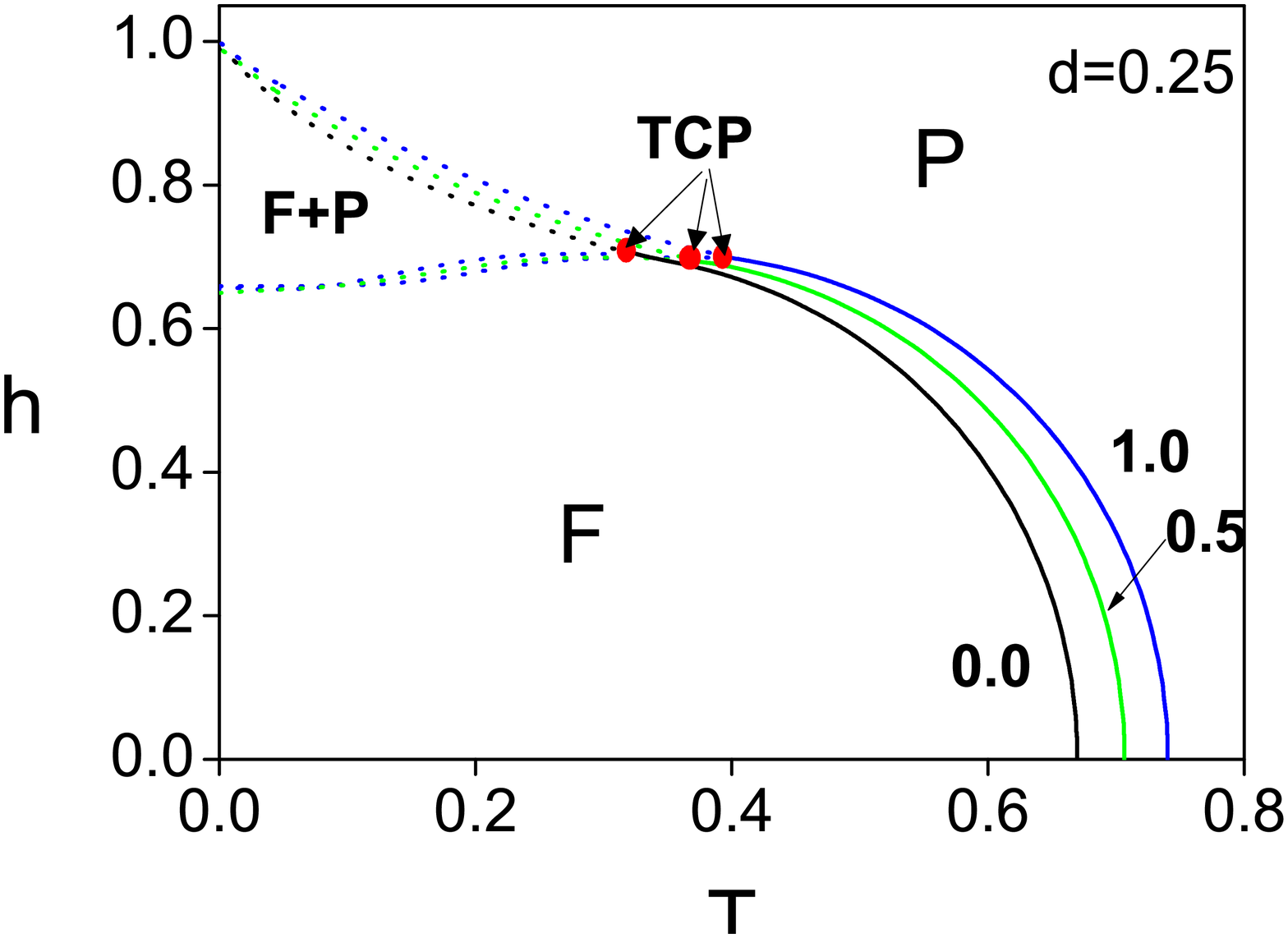}}
\subfigure[\hspace{0.4cm}]{\label{fig:sub:b}
\includegraphics[width=6.0cm,height=6.0cm,angle=0]{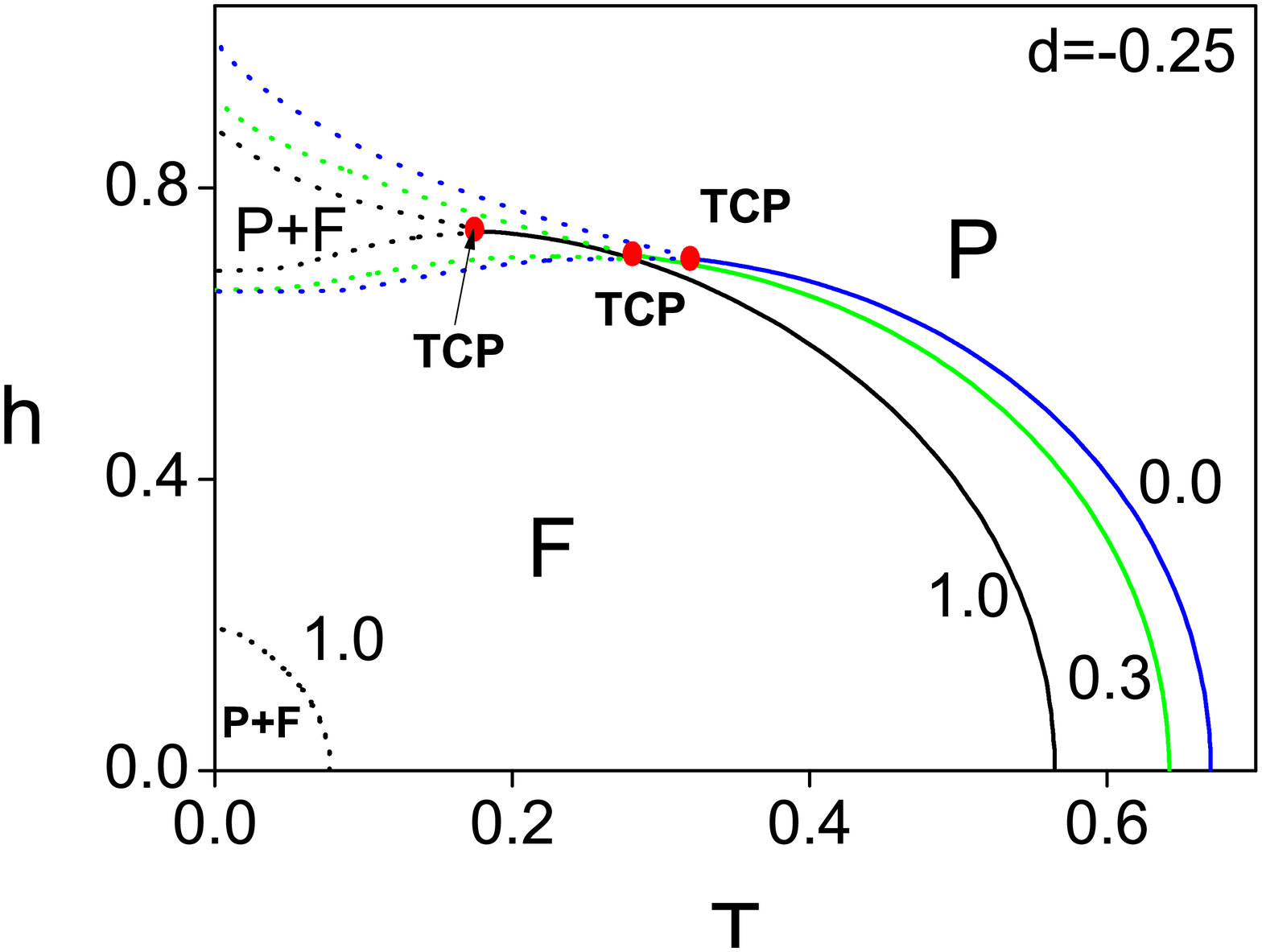}}
\subfigure[\hspace{0.4cm}]{\label{fig:sub:a}
\includegraphics[width=6.0cm,height=6.0cm,angle=0]{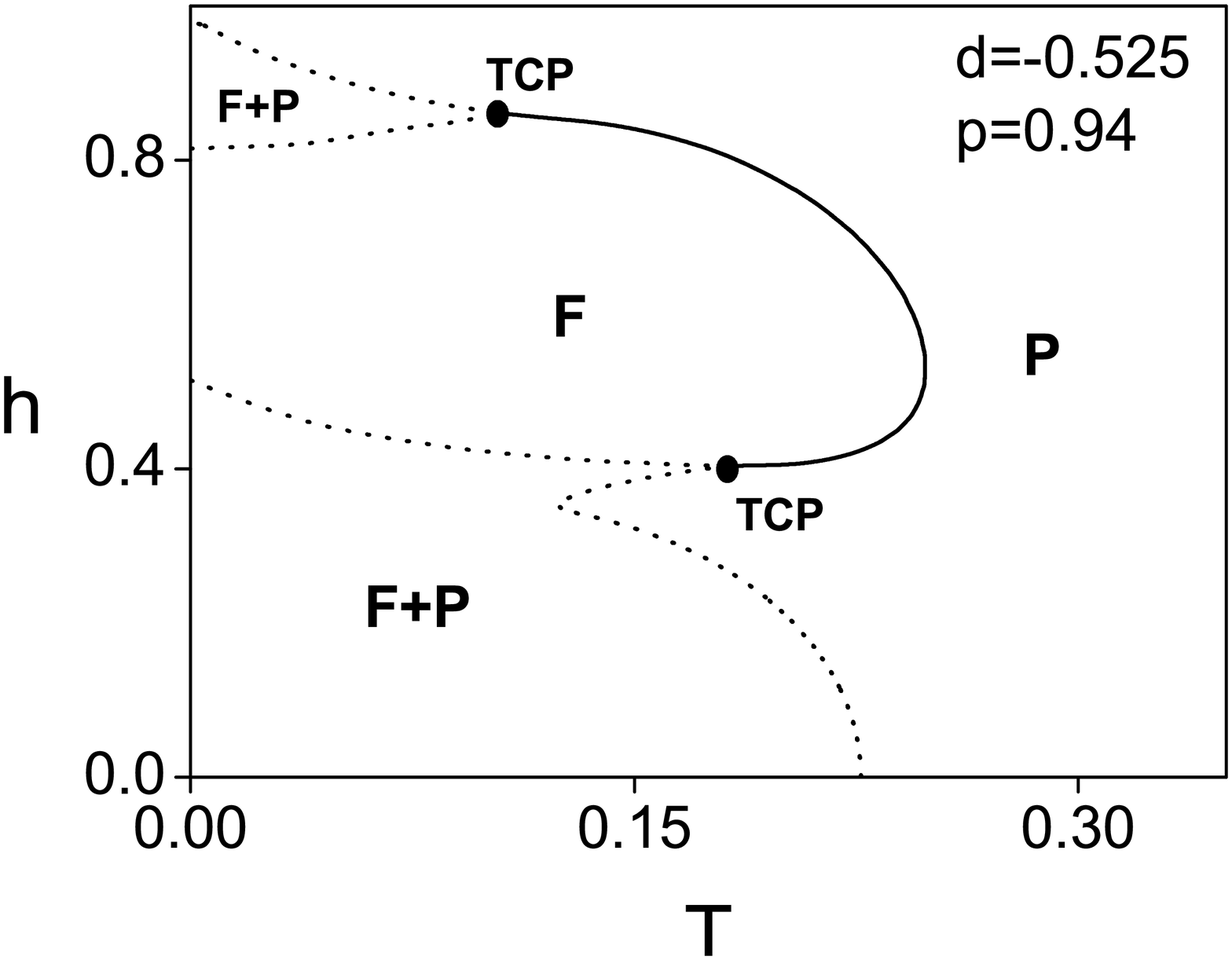}}
\subfigure[\hspace{0.4cm}]{\label{fig:sub:b}
\includegraphics[width=6.0cm,height=6.0cm,angle=0]{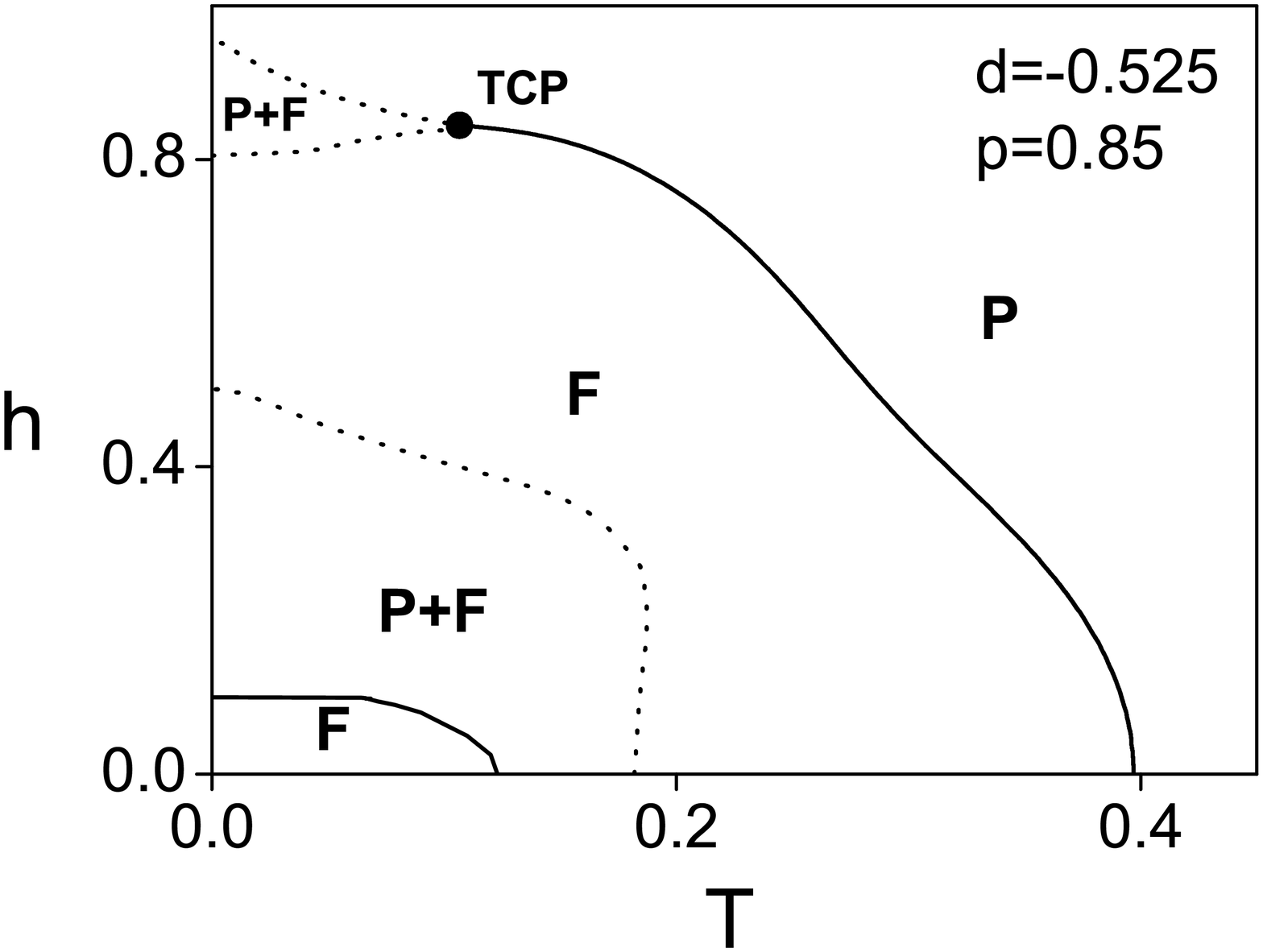}}
\subfigure[\hspace{0.4cm}]{\label{fig:sub:b}
\includegraphics[width=6.0cm,height=6.0cm,angle=0]{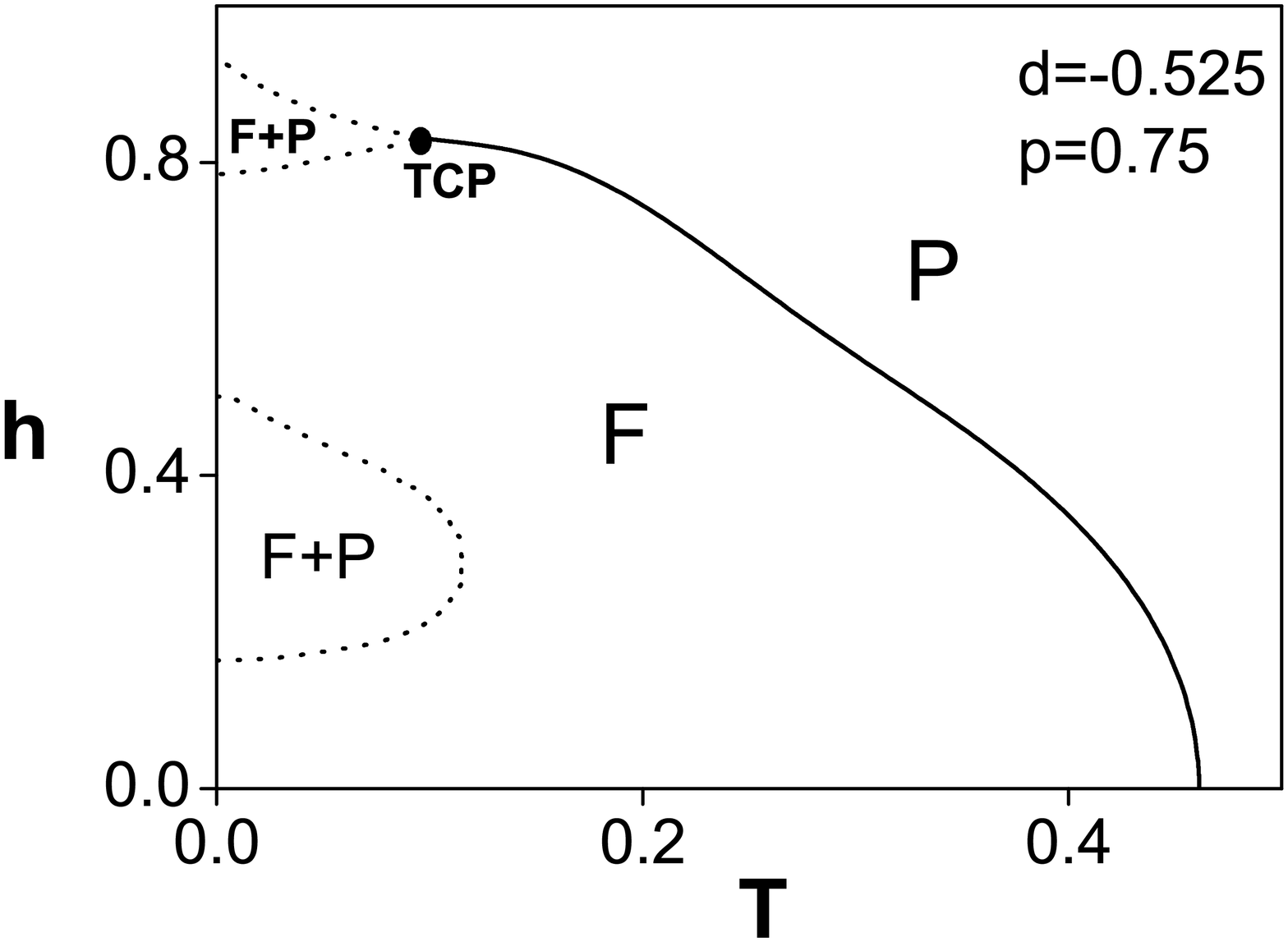}}
\subfigure[\hspace{0.4cm}]{\label{fig:sub:c}
\includegraphics[width=6.0cm,height=6.0cm,angle=0]{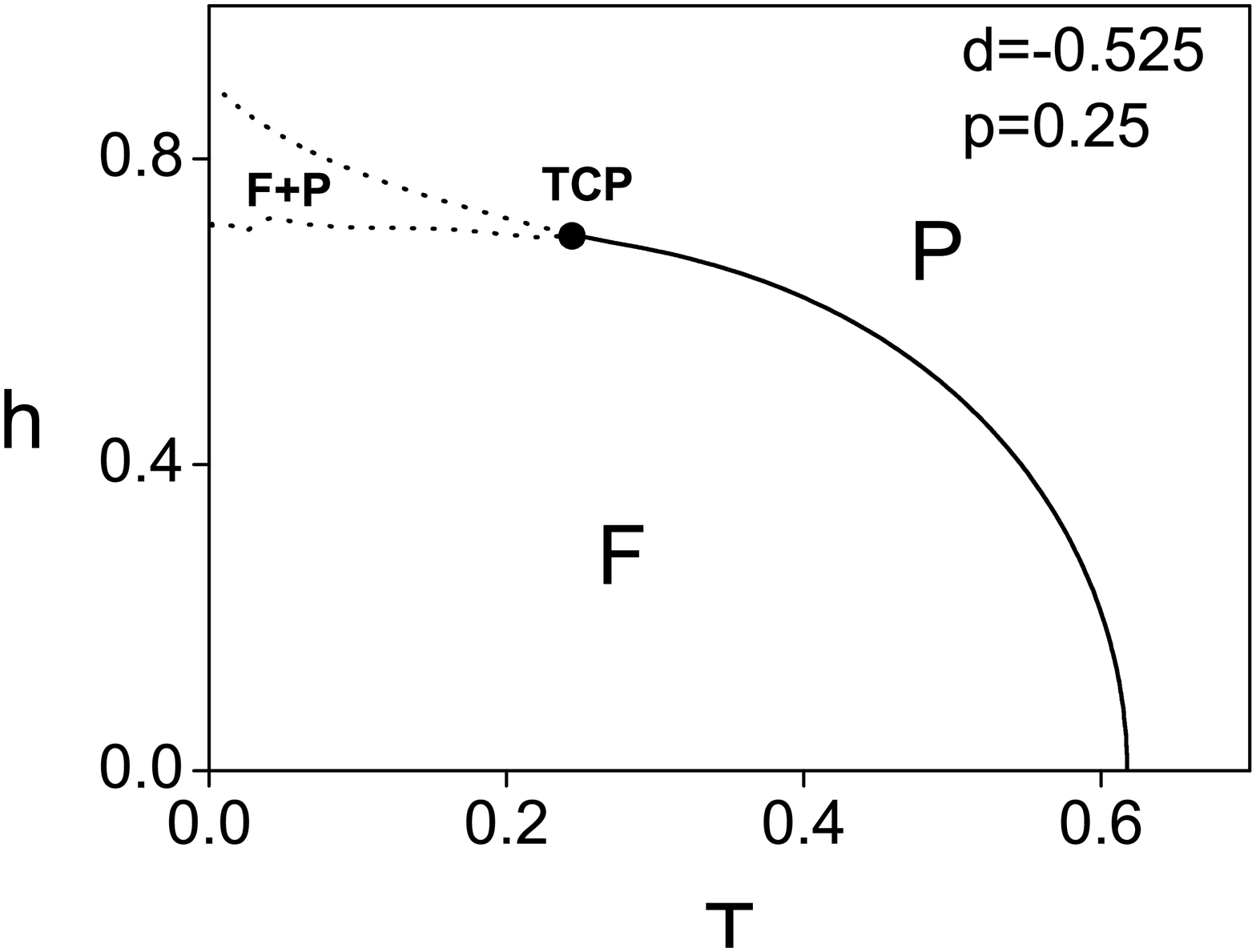}}
\end{center}

\caption{Dynamic phase diagrams of the Blume-Capel model with crystal field randomness  in the (T,h)
plane for various values of the single ion anisotropy concentration (p). Dotted and
solid lines represent the first-order and
second-order phase transitions, respectively. (a) d=0.25, the number accompanying each curve
denotes the value of p. (b) d=-0.25  the number accompanying each curve illustrates the value of p.
(c) d=-0.525 and p=0.94. (d) d=-0.525 and p=0.85. (e) d=-0.525 and p=0.75. (f) d=-0.525 and p=0.25.
} \label{fig:sub:Fig2a-Fig2d}
\end{figure}
\begin{figure}[tbp]
\begin{center}
\subfigure[\hspace{0.4cm}]{\label{fig:sub:a}
\includegraphics[width=6.0cm,height=6.0cm,angle=0]{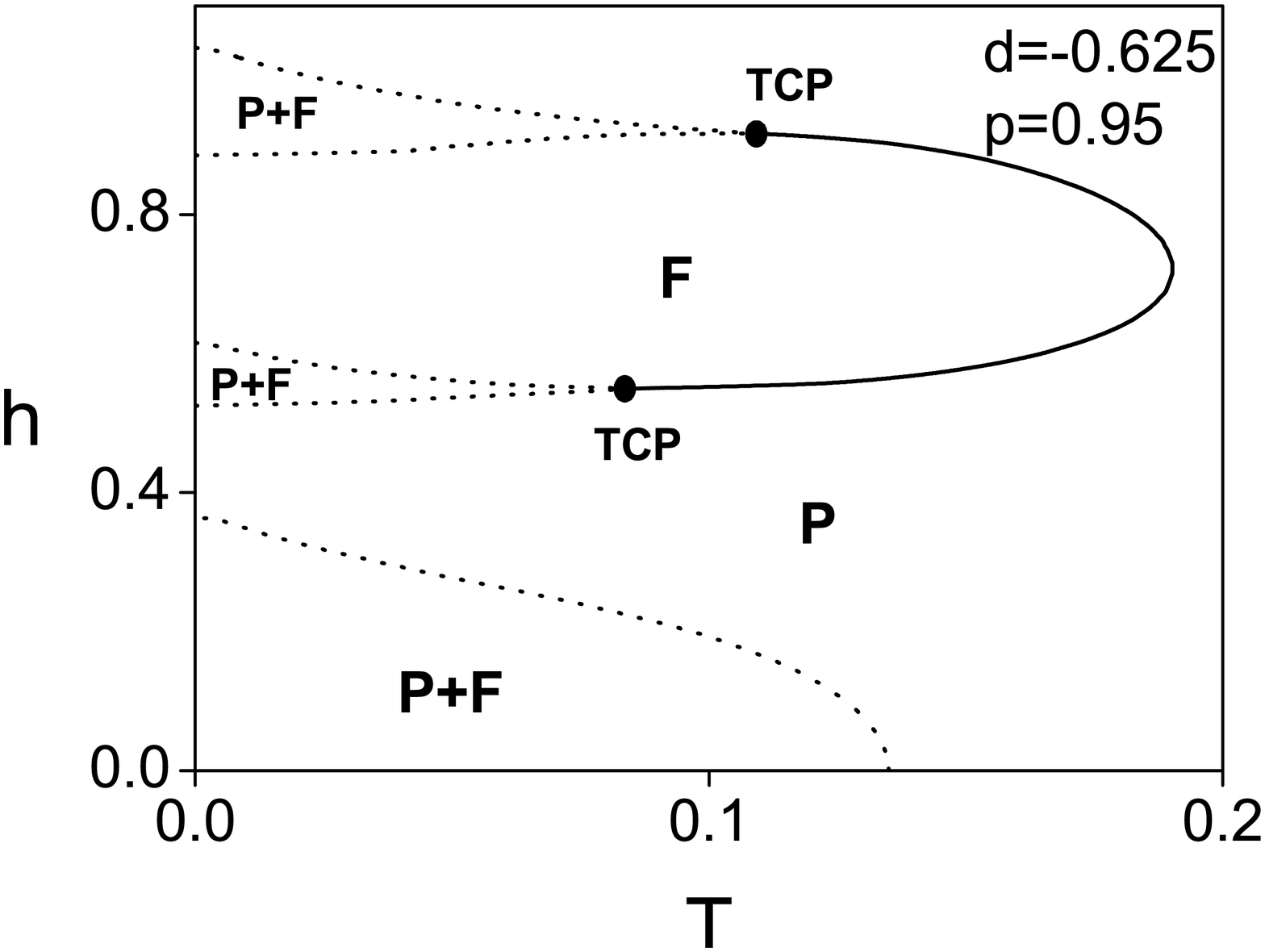}}
\subfigure[\hspace{0.4cm}]{\label{fig:sub:b}
\includegraphics[width=6.0cm,height=6.0cm,angle=0]{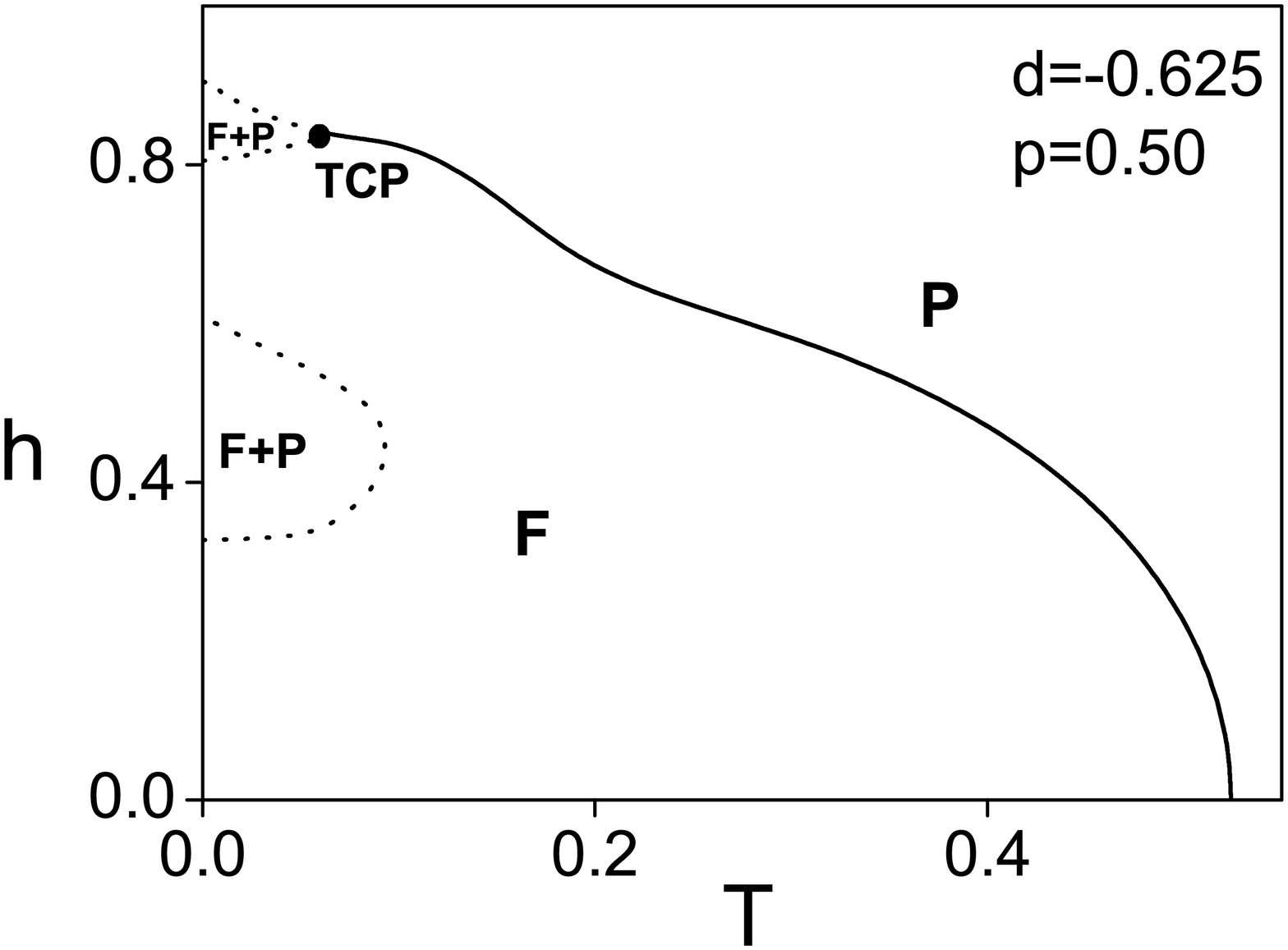}}
\subfigure[\hspace{0.4cm}]{\label{fig:sub:a}
\includegraphics[width=6.0cm,height=6.0cm,angle=0]{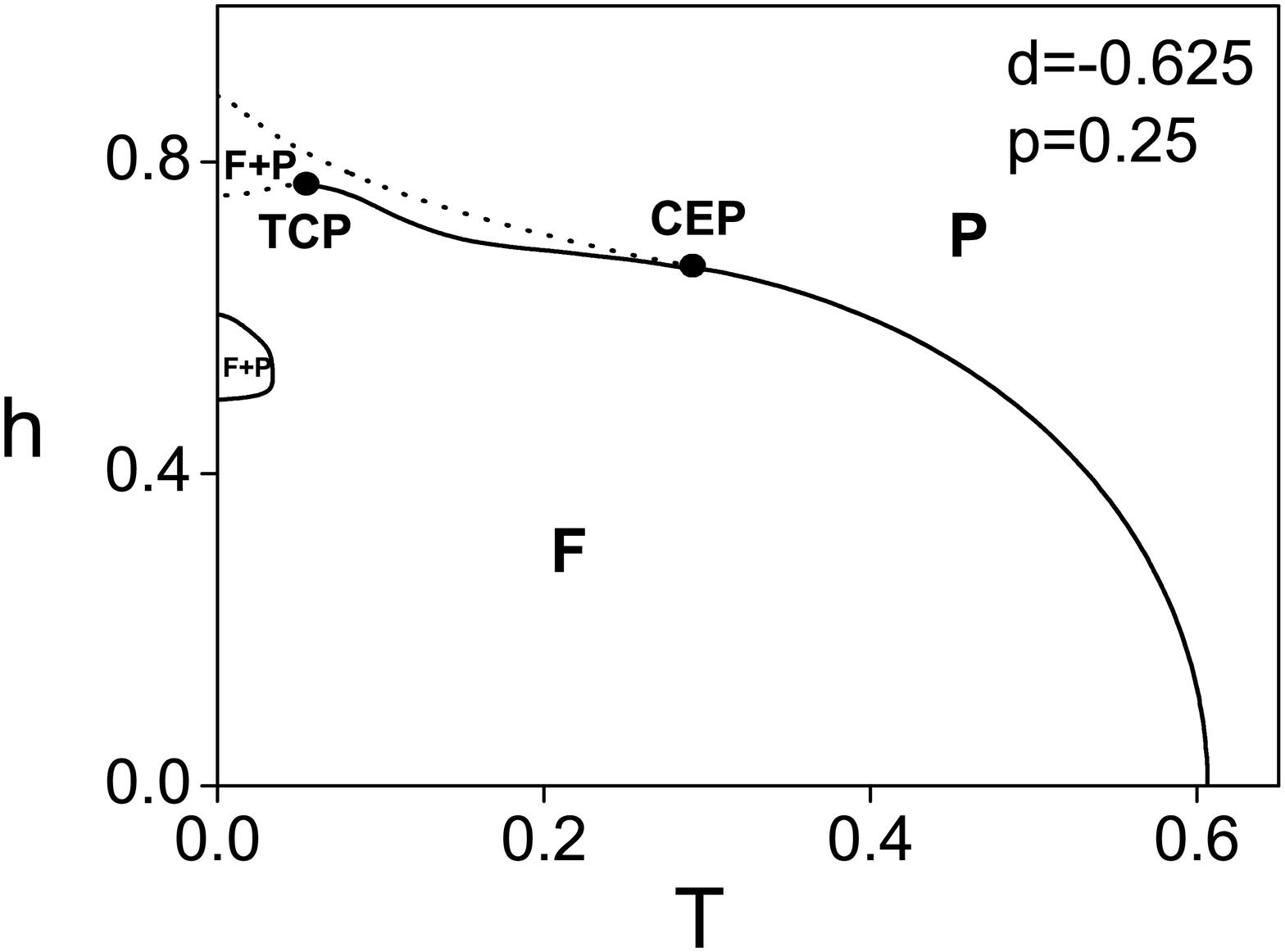}}
\subfigure[\hspace{0.4cm}]{\label{fig:sub:b}
\includegraphics[width=6.0cm,height=6.0cm,angle=0]{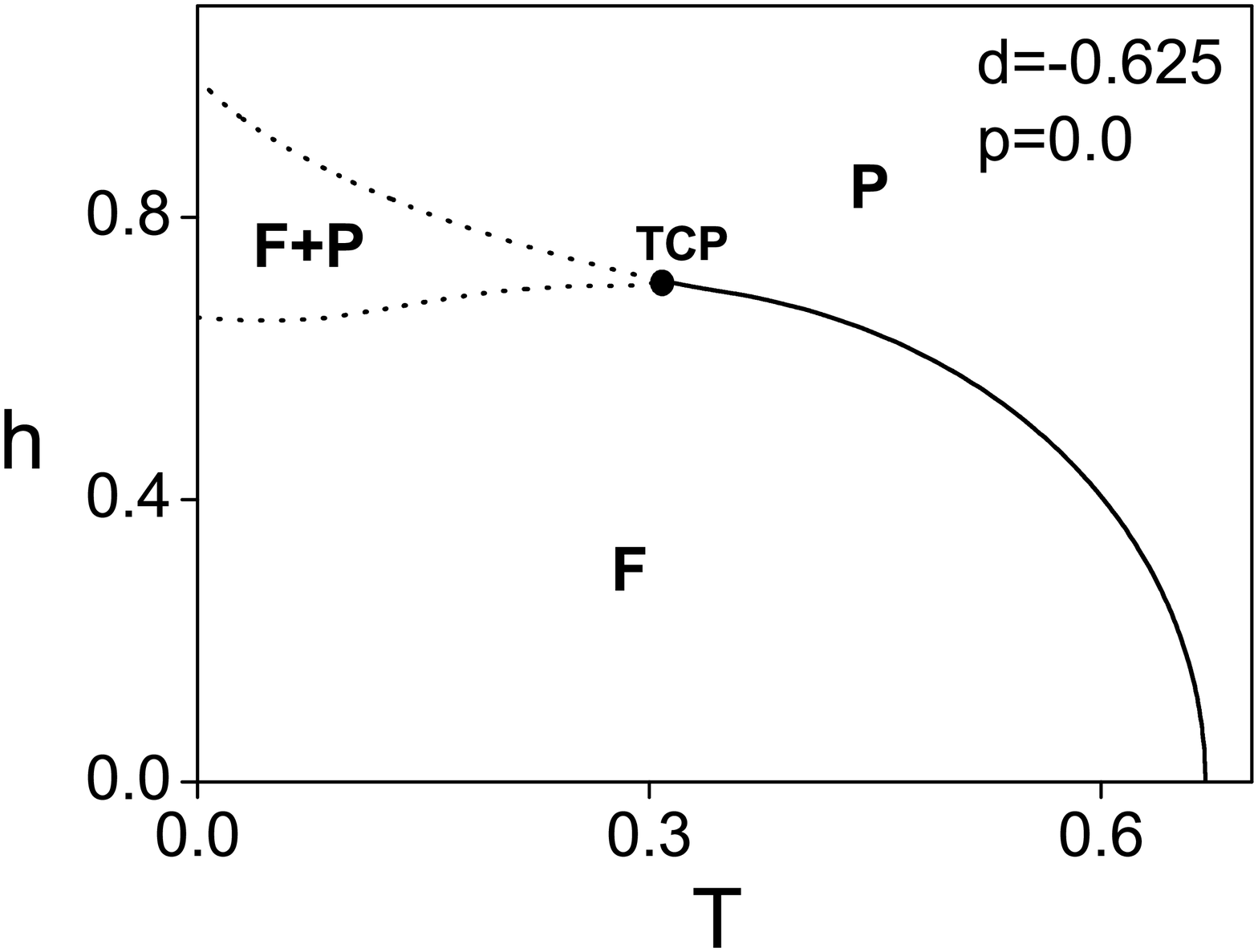}}
\end{center}

\caption{Dynamic phase diagrams of the Blume-Capel model with crystal field randomness in the (T,h)
plane for various values of the single ion anisotropy concentration (p) while d=-0.625. Dotted and
solid lines represent the first-order and second-order phase transitions, respectively. (a) p=0.95,
(b) p=0.50, (c) p=0.25, (d) p=0.0.
} \label{fig:sub:Fig2a-Fig2d}
\end{figure}

\begin{figure}[tbp]
\begin{center}
\subfigure[\hspace{0.4cm}]{\label{fig:sub:a}
\includegraphics[width=6.0cm,height=6.0cm,angle=0]{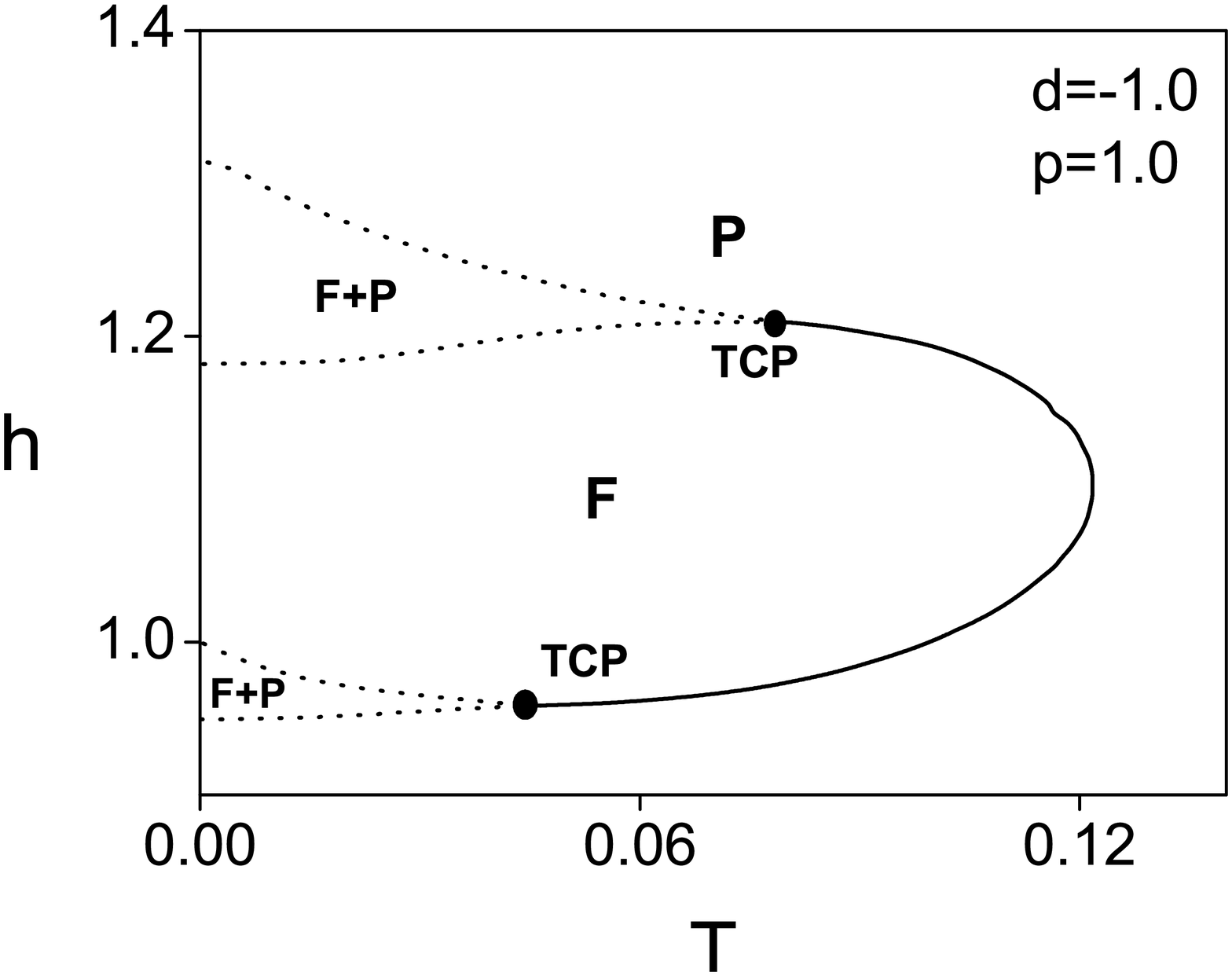}}
\subfigure[\hspace{0.4cm}]{\label{fig:sub:b}
\includegraphics[width=6.0cm,height=6.0cm,angle=0]{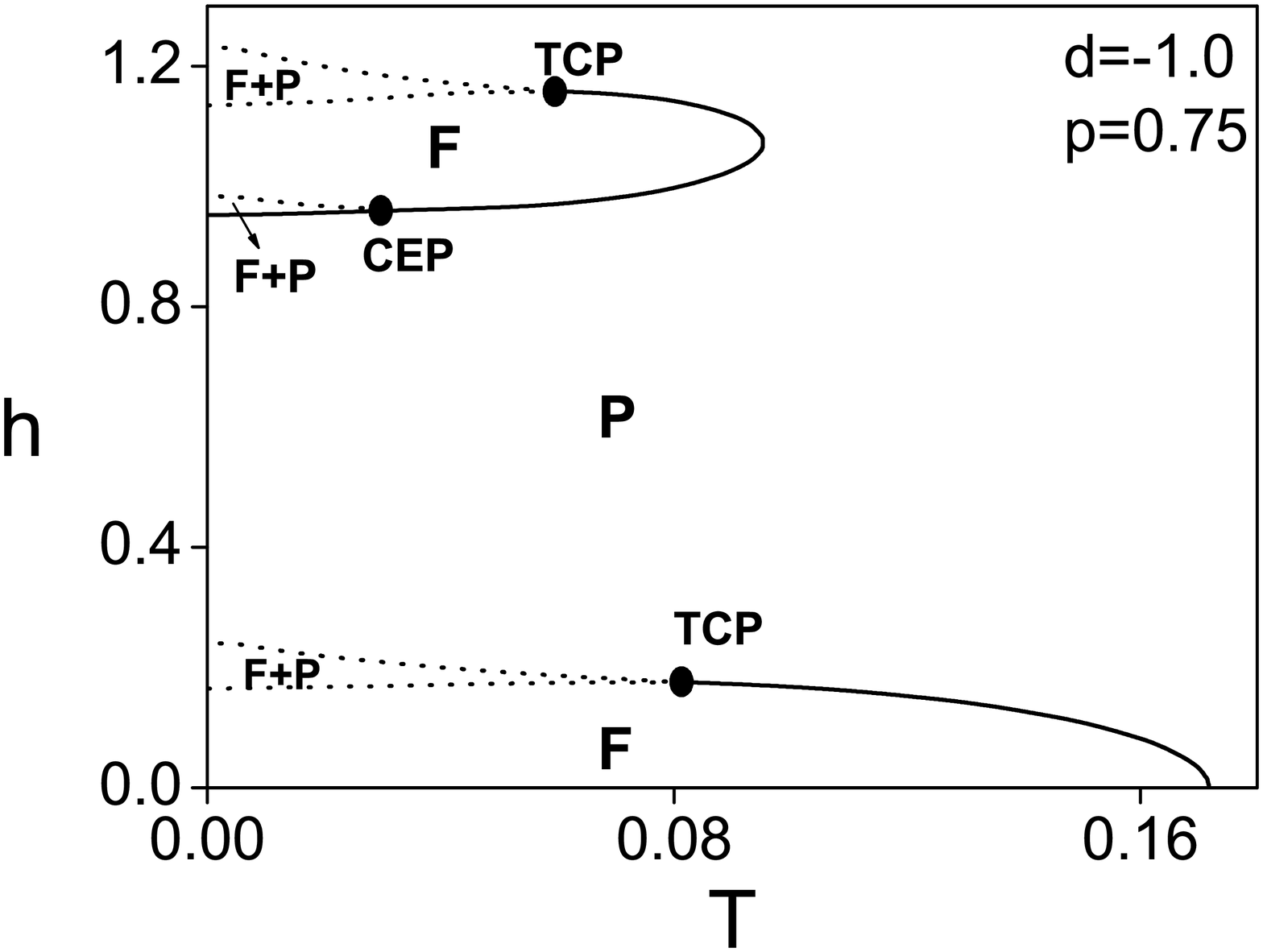}}
\subfigure[\hspace{0.4cm}]{\label{fig:sub:a}
\includegraphics[width=6.0cm,height=6.0cm,angle=0]{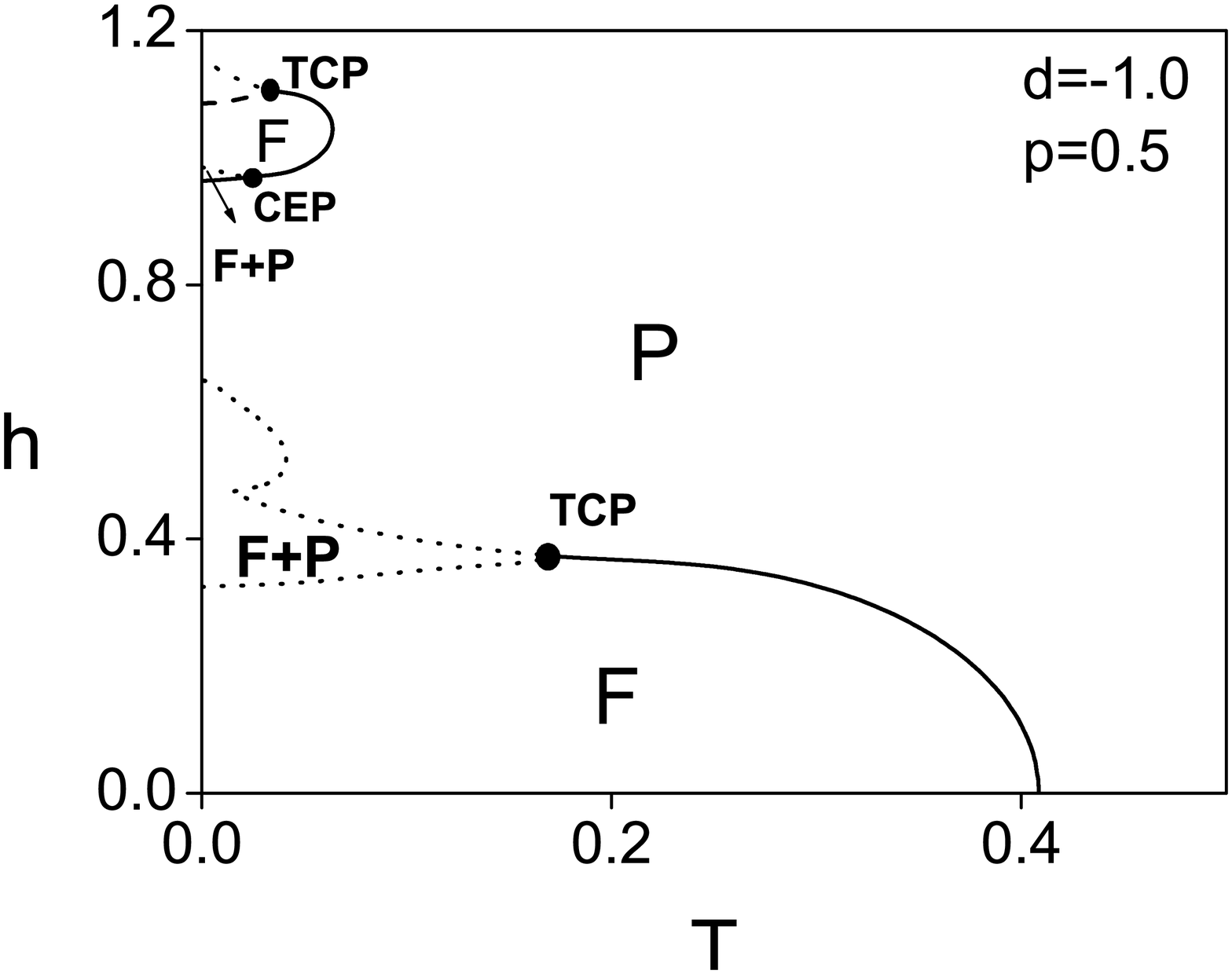}}
\subfigure[\hspace{0.4cm}]{\label{fig:sub:b}
\includegraphics[width=6.0cm,height=6.0cm,angle=0]{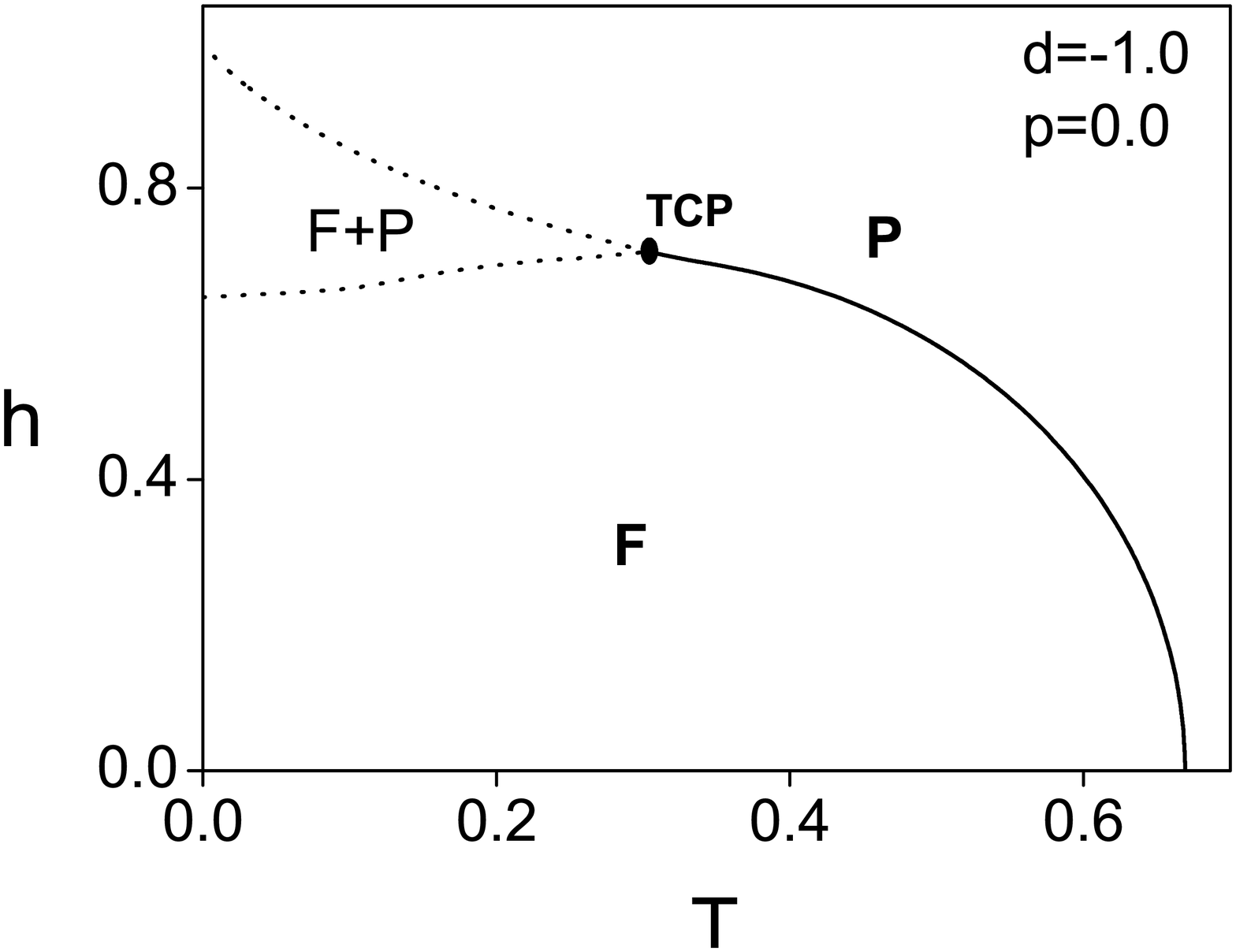}}
\end{center}

\caption{Dynamic phase diagrams of the Blume-Capel model with crystal field randomness in the (T,h)
plane for different values of the single ion anisotropy concentration (p) while d=-1.0. Dotted and
solid lines represent the first-order and second-order phase transitions, respectively. (a) p=1.0,
(b) p=0.75, (c) p=0.5, (d) p=0.0.
} \label{fig:sub:Fig2a-Fig2d}
\end{figure}

 \end{document}